\documentclass{aa}

\usepackage{natbib}
\usepackage{graphicx}
\usepackage{booktabs}
\usepackage{placeins}
\usepackage[varg]{txfonts}
\usepackage{stfloats} 
\usepackage{siunitx}
\usepackage{csquotes}
\usepackage{threeparttable}
\usepackage[inline]{enumitem}   
\usepackage{orcidlink}
\usepackage{indentfirst}

\bibpunct{(}{)}{;}{a}{}{,} 
\AddToHook{begindocument/before}{\usepackage{hyperref}}

\begin{document}

\title{Activity of main-belt comet 324P/La Sagra}
\author
{\textbf{M. Mastropietro\inst{1,}\inst{2,} \orcidlink{0000-0002-4966-4639} 
, Y. Kim\inst{1,}\inst{3} \orcidlink{0000-0002-4676-2196}
, H. H. Hsieh\inst{4} \orcidlink{0000-0001-7225-9271} 
, J. Agarwal\inst{1,}\inst{2} \orcidlink{0000-0001-6608-1489} 
}}
\institute{Institut für Geophysik und Extraterrestrische Physik, Technische Universität Braunschweig, Mendelssohnstraße 3, 38106 Braunschweig, Germany; \email{m.mastropietro@tu-braunschweig.de}
\and Max Planck Institute for Solar System Research, Justus-von-Liebig-Weg 3, 37077 Göttingen, Germany 
\and Department of Earth, Planetary and Space Sciences, UCLA, Los Angeles, CA 90095-1567, USA
\and Planetary Science Institute, 1700 East Fort Lowell Road, Suite 106, Tucson, 85719, USA}

\date{Received 12 June 2024 / 
Accepted 5 September 2024}

\authorrunning{Mastropietro et al., 2024}

\abstract{}{We study the activity evolution of the main-belt comet 324P/La Sagra over time and the properties of its emitted dust.}{We performed aperture photometry on images taken by a wide range of telescopes at optical and thermal infrared wavelengths between 2010 and 2021. We derived the combined scattering cross section of the nucleus and dust (when present) as a function of time, and we derived the thermal emission properties.}{Fitting an IAU H-G phase function to the data obtained when 324P was likely inactive, we derived an absolute nucleus magnitude $H_R = (18.4 \pm 0.5)$ mag using $G = 0.15 \pm 0.12$. The activity of 324P/La Sagra during the 2015 perihelion passage has significantly decreased compared to the previous perihelion passage in 2010, and it decreased even further during the 2021 perihelion passage. This decrease in activity may be attributed to mantling or to the depletion of volatile substances.
The $Af\rho$ profile analysis of the coma of the main-belt comet suggests a near-perihelion transition from a lower-activity pre-perihelion to a higher-activity post-perihelion steady state.
We calculate a  dust geometric albedo in the range of (2 - 45)\%, which prevents us from constraining the spectral type of 324P/La Sagra, but we found an indication of dust superheating at 4.5 µm.}{}

\maketitle

\renewcommand{\thefootnote}{\arabic{footnote}}

\section{Introduction}
Active asteroids exhibit comet-like dust emission, but have asteroid-like orbits (with a Tisserand parameter with respect to Jupiter $T_{J} \gtrsim 3$, and with a location inside the Jupiter orbit). Main-belt comets (MBCs) are a subgroup of active asteroids \citep{Jewitt2015, HsiehSheppard2015}. The dust emission in MBCs is due to ice sublimation \citep{Hsieh2012a, Jewitt2015}, while in active asteroids that are not driven by sublimation, it is due to events such as impacts \citep{Jewitt2011, Bodewits2011, Ishiguro2011, Kim2017a, Kim2017b} or rotational destabilization \citep{Jewitt2013, Jewitt2014a, Drahus2015, Sheppard2015}.
Observationally, MBCs are usually identified through the specific time-dependence of the activity: They are active near perihelion, and the activity continues for an extended period of time of at least a few weeks, and it recurs during successive perihelia. In non-MBC active asteroids, the activity can occur at any true anomaly and can consist of single-time events, but is not limited to them.

Currently, only the \textit{James Webb} Space Telescope (JWST) has a sufficiently high sensitivity to spectroscopically detect water vapor near an MBC. The first such detection was reported for 238P/Read \citep{Kelley2023}. This was the first detection of water-outgassing from a main-belt object that also emits visible dust. \citet{Kuppers2014} detected water-vapor plumes without associated dust from the dwarf planet Ceres.

In the absence of spectroscopic evidence, repeated and prolonged dust-emission activity in MBCs near perihelion is still considered a strong indicator of sublimation, because this behavior is difficult to explain as the direct result of other mechanisms \citep{Hsieh2008, Hsieh2011, Hsieh2012a, Hsieh2012b, HH2015, Moreno2011, Moreno2013, Jewitt2014b, Pozuelos2015}.
However, the survival of primordial water ice in main-belt objects after billion-year timescales has been shown to be possible by numerical thermal models only when the ice is buried under a protective dusty layer \citep{Fanale1989, Schorghofer2008, Schorghofer2016, Prialnik2009, SchorghoferHsieh2018}. Hence, MBC activity needs a trigger event to expose this buried ice to solar irradiation. One possible trigger may be collisions \citep{Haghighipour2016, Haghighipour2018}. The dust emission of the MBCs is then driven by the sublimation of these recently exposed materials \citep{Hsieh2004, Hsieh2006}. Direct ice exposure is not necessary to activate MBCs, because a mere reduction of the dust layer thickness at the bottom of a crater may be sufficient for the underlying ice to sublimate \citep{Capria2012}. 
Most MBCs have their peak activity after perihelion. This delay in the sublimation process may be due to the time needed for the thermal wave to reach the ice buried in the subsurface \citep{HH2015}.

However, fast rotation may also play a role in triggering ice sublimation and/or sustaining dust emission against gravity.
For example, the activity of 133P/Elst-Pizarro, the first MBC discovered in 1996, may be due to the combined effects of an initial triggering impact, sublimation, and rapid rotation \citep{Hsieh2004, Hsieh2010, HH2015, Jewitt2014b}. 

Numerical models show that the orbits of most MBCs are dynamically stable, indicating that these objects formed in the main asteroid belt \citep{Haghighipour2009, Jewitt2009, Hsieh2012a, Hsieh2012b}. There they would have been dormant for a long time until their recent activation \citep{Hsieh2004, Capria2012}. Some studies also suggested that MBC activation may be facilitated by the preceding collisional breakup of larger parent bodies that would leave subsurface ice at comparatively shallow depths \citep{Hsieh2018b}. However, the orbits of a few MBCs are unstable on timescales of 20-30 Myr, suggesting that they may have reached their current orbital locations through interactions with giant planets \citep{Haghighipour2009, Jewitt2009}. Some MBCs may even have formed in the outer Solar System and been captured into the main asteroid belt \citep{Haghighipour2009, Hsieh2016, kim-agarwal2022}.

The MBC 324P/La Sagra (hereafter 324P) was discovered in 2010 as P/2010 R2 \citep{Nomen2010}. Its prolonged dust emission and mass loss during different perihelion passages suggest that its activity is driven by sublimation \citep{Moreno2011, Hsieh2012b, Bauer2012, HsiehSheppard2015}. It is dynamically associated with the Alauda family \citep{Hsieh2018b}, meaning that the composition of 324P may be similar to that of the other Alauda family asteroids. The orbital elements of 324P are listed in Table~\ref{table:2}.

In this work, we perform a photometric analysis of 324P during its 2010, 2015 and 2021 perihelion passages, and we compare our results to previous studies. In Sect.~\ref{sec:data}, we describe the image datasets used for this work. In Sect.~\ref{sec:results}, we report the methods and analyses applied to the image sets and present and discuss our results. We summarize the results in Sect.~\ref{sec:summary}.

\begin{table}[t]
\begin{threeparttable}
\caption{Parameters describing the orbit of 324P. }
\label{table:2}
\begin{tabular}{lcc}
\toprule
Parameter& Value&Unit\\
\midrule
Eccentricity ($e$) & 0.154 &  \\
Semimajor axis ($a$)  & 3.095 & AU \\
Perihelion distance ($q$)  & 2.619 & AU \\
Aphelion distance ($Q$)  & 3.570 & AU \\
Inclination ($i$)  & 21.402 & deg \\
Orbital period ($P$)  & 5.448 & yr\\
Tisserand parameter ($T_J$) & 3.100 & \\
\bottomrule
\end{tabular}
\begin{tablenotes}\small{
\item Note: All quantities except the Tisserand parameter were obtained from the JPL HORIZONS System for the epoch JD 2460053.5 (2023-04-19) Barycentric Dynamical Time (TDB).
}
\end{tablenotes}
\end{threeparttable}
\end{table}

\begin{figure}[t]
\resizebox{\hsize}{!}{\includegraphics{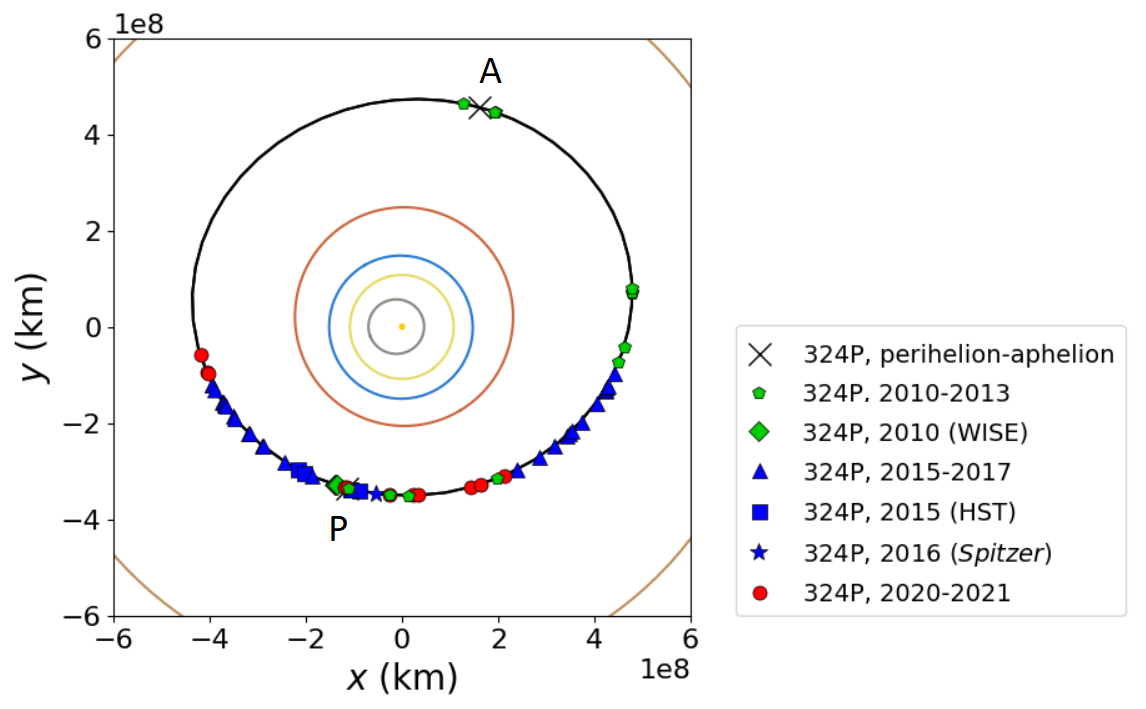}}
\caption{Orbit plot in the heliocentric ecliptic coordinate system. The origin is the center of the Sun, the plane of reference is the ecliptic plane, and the direction of the x-axis points toward the vernal equinox. The plot shows the positions of 324P at the epochs of our observations with the orbits of Mercury, Venus, Earth, Mars, 324P, and Jupiter. Crosses represent the perihelion (P) and aphelion (A) positions. Pentagons, triangles, squares and circles represent visual data, and diamonds and stars are IR data. (The plot was generated using the poliastro Python library \citep{poliastro}.) }
\label{orbit}
\end{figure}

\section{Observations} 
\label{sec:data}
We studied archival data of 324P obtained between 2010 and 2020, and we observed it during its perihelion passage in 2021 (Fig.~\ref{orbit}). 
324P was at perihelion on 2010 June 25, 2015 November 30, and 2021 May 6 (UT). 

Images captured at both optical and infrared (IR) wavelengths were identified using the Canadian Astronomy Data Centre (CADC) website\footnote{\url{https://www.cadc-ccda.hia-iha.nrc-cnrc.gc.ca/en/ssois/index.html}.} \citep{Gwyn2012} and from the Infrared Science Archive (IRSA) website\footnote{\url{https://irsa.ipac.caltech.edu/frontpage/}.}. 
All visible-light images used in our study were obtained in the R band, except for the Pan-STARRS (PS1) image dated 2010 June 26 (which is in the z band), four PS1 images from 2010 September 8 (g band), and \textit{Hubble} Space Telescope (HST) images in F350LP filter, whose pivot wavelength \citep{WFCHandbook} is at 587.39 nm, similar to the V band.

Images obtained with the Very Large Telescope (VLT) on 2019 April 12 have a signal-to-noise ratio of S/N $<$ 8, and we therefore excluded them from the further analysis. We did not detect 324P in \textit{Zwicky} Transient Facility (ZTF) images from 2019 April 9, 2019 April 12, 2019 April 19, and 2019 April 28.

\subsection{Visible-light data}
Visible-light images of 324P (Fig.~\ref{mosaic}) were obtained with the telescopes and instruments listed in Table~\ref{table:1}.
We performed bias subtraction and flat-fielding using the following software packages: Dragons for Gemini data, Image Reduction and Analysis Facility (\citealp[IRAF;][]{iraf1, iraf2}) to process New Technology Telescope (NTT) data, and EsoRex for VLT data. The Pan-STARRS data were processed by the PS1 Image Processing Pipeline \citep{Magnier2006}. Isaac Newton Telescope (INT) data were processed by the INT Wide Field Camera pipeline \citep{Irwin2001}. Canada-France-Hawaii Telescope (CFHT) data were processed by the Elixir pipeline \citep{Magnier2004}. Lowell Discovery Telescope (LDT) data were received in private communication from Matthew M. Knight in a post-processed state. For the HST data, we computed the target magnitude using the relation
\begin{equation} \label{eq:3}
V = 20 - 2.5 \log_{10} \left[\frac{f}{n} \right],
\end{equation}
which we obtained from the Exposure Time Calculator (ETC) for the Wide Field Camera 3 (WFC3) Ultraviolet-Visible (UVIS) channel\footnote{\url{https://etc.stsci.edu/etc/input/wfc3uvis/imaging/}.}. 
$V$ is the apparent magnitude in the V band, $f$ is the count rate of the source in e$^-$ s$^{-1}$, and $n$ is the count rate\footnote{ETC Request IDs: WFC3UVIS.im.1845086, WFC3UVIS.im.1845087, WFC3UVIS.im.1845088, WFC3UVIS.im.1845089.} obtained by the ETC with the F350LP filter from a source with a Sun-like (Kurucz G2V) spectrum renormalized to Vega magnitude 20 in the Johnson/V filter. 

For the CFHT and PS1 images, we used zero points computed by the Elixir and PS1 IPP pipelines, respectively. 
For data from all other instruments, we performed a photometric calibration relative to field stars from the PS1 catalog\footnote{\url{https://ps1images.stsci.edu/cgi-bin/ps1cutouts}.}.
We also applied this method to the CFHT and PS1 data to check for consistency within our dataset, which we confirmed.
We assumed solar colors for the conversions between the different PS1 bands $r-z=0.5$ and $r-g=-0.62$ \citep{Tonry2012, Willmer2018}, and for the conversion from the V to R band, $V-R=0.35$ \citep{Holmberg2006, Jewitt2016}.

With IRAF, we performed photometry on single images of 324P using a circular aperture with a fixed physical radius of 2300 km at the comet. We opted for a constant physical radius (as opposed to constant angular size) to capture the same volume around the nucleus in every measurement. 
Subsequently, we calculated the weighted average of the apparent magnitudes measured on single exposures for each dataset, resulting in one measurement point per epoch and telescope (Table~\ref{table:1}).
We performed circular aperture photometry on stacked HST images \citep{Jewitt2016} with a fixed physical radius of 3000 km at the comet to ensure consistency in our comparative analysis with the \textit{Spitzer} data (Section~\ref{IRdata}). 

\begin{figure}[t]
\centering
\resizebox{\hsize}{!}{\includegraphics{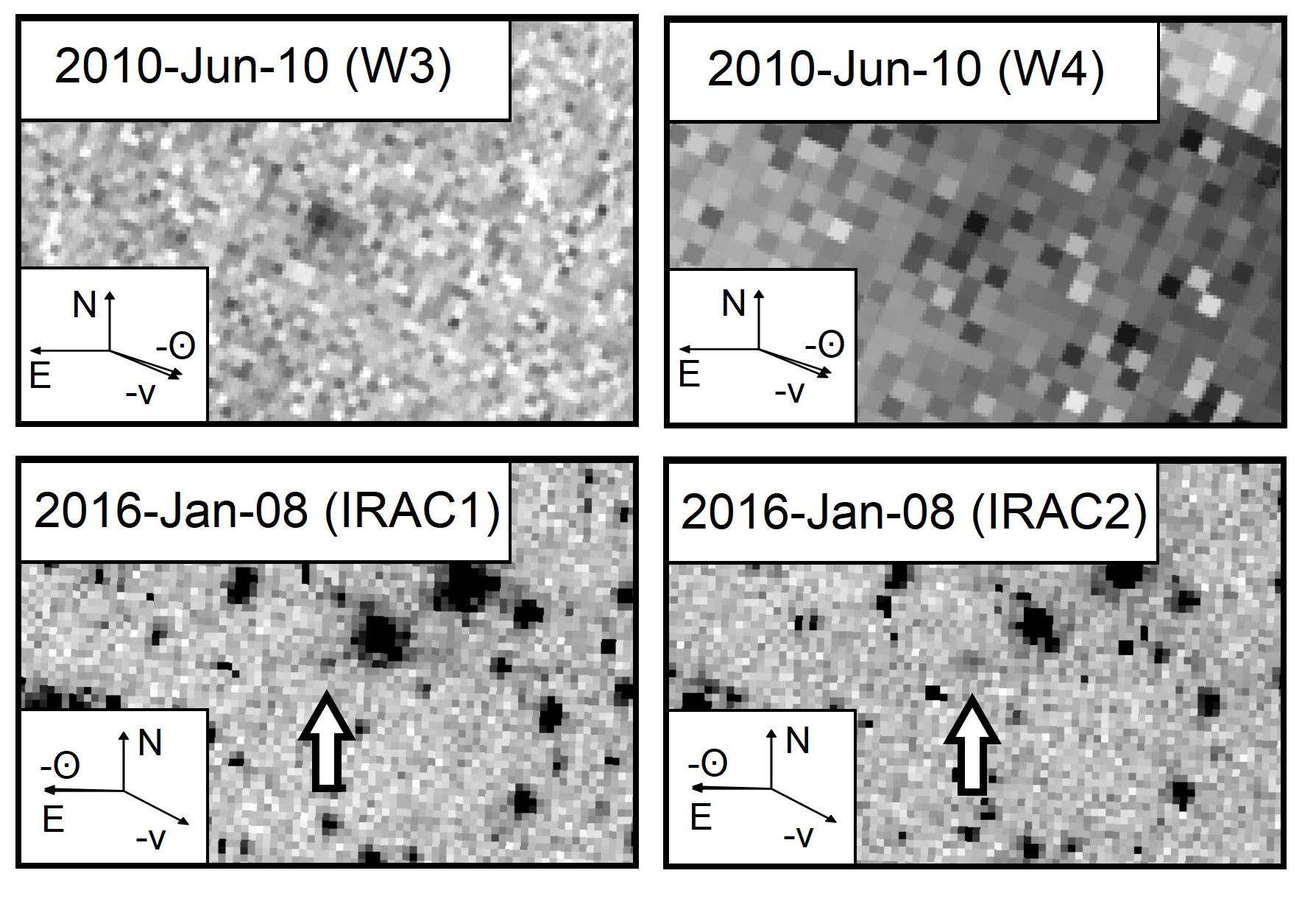}}
\caption{IR images of 324P (at the center of each panel). The label W3 refers to WISE band 3 (12 µm), W4 to WISE band 4 (22 µm), and IRAC1 and IRAC2 corresponds to \textit{Spitzer} Space Telescope observations at 3.6 µm and 4.5 µm, respectively. The dates and directions on sky are labeled as in Fig.~\ref{mosaic}.}
\label{WISESPITZER}
\end{figure}

\subsection{Infrared data}\label{IRdata}
Infrared images of 324P (Fig.~\ref{WISESPITZER}) were taken on 2010 June 9 -- 11 (true anomaly of $\nu$ = -3.9°, in the active phase) with the Wide-field Infrared Survey Explorer (WISE) telescope with a 0.4-meter diameter primary mirror (\SI{2.75}{\arcsecond} per pixel in the W1 band at 3.4 µm, the W2 band at 4.6 µm, and the W3 band at 12 µm, \SI{5.5}{\arcsecond} per pixel in the W4 band at 22 µm). 
The WISE data were processed and photometrically calibrated by the WSDS PIPELINES \citep{Cutri2012}. To increase the S/N, we co-added images with the WISE Coadder software\footnote{\url{https://irsa.ipac.caltech.edu/applications/ICORE/}.}. 

We converted WISE magnitudes into fluxes using \citep{Cutri2012} 
\begin{equation} \label{eq:5}
  F_{\nu}=   F_{\nu0} \cdot 10^{(-m_{Vega}/2.5)},
\end{equation}
where $F_{\nu}$ is the flux density in Jy, $F_{\nu0}$ is the zero-magnitude flux density \citep{Wright2010, Mainzer2011, Masiero2011}, and $m_{Vega}$ is the measured WISE Vega magnitude, all at frequency $\nu$. 
We measured the dust brightness inside apertures with angular radii \SI{11}{\arcsecond} (corresponding to 19230 km) for the W1, W2 and W3 bands, and \SI{22}{\arcsecond} (38460 km) for the W4 band \citep{Bauer2012}. We chose aperture values that were sufficiently larger than the point spread function (PSF) full width at half maximum (FWHM) of each band. For bands W1 and W2, we were only able to derive upper limits. Our results for all four bands are consistent with those of \citet{Bauer2012}.

Infrared images were also taken on 2016 January 8 ($\nu$ = 9.8°, also in the active phase) with the Infrared Array Camera (IRAC) on the \textit{Spitzer} Space Telescope with a 0.85-meter diameter primary mirror (\SI{1.22}{\arcsecond} \si{pixel^{-1}} in channels 1 (3.6 µm) and 2 (4.5 µm)). \textit{Spitzer} data were processed and calibrated by the IRAC pipeline \citep{Fazio2004, IRAC_handbook}. 

\begin{figure}[t]
\centering
\resizebox{\hsize}{!}{\includegraphics{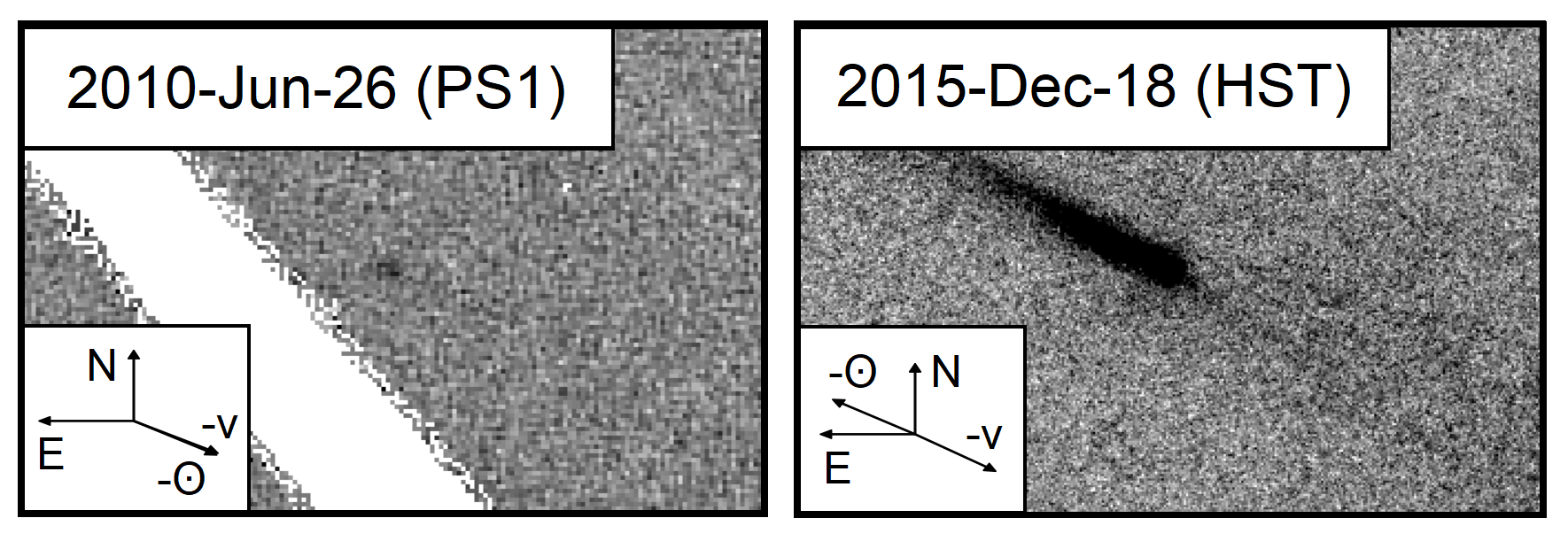}}
\caption{Visual-light images (PS1 in the z band and HST in the V band) of 324P (at the center of each panel). The dates and directions on sky are labeled as in Fig.~\ref{mosaic}. These visual images were obtained closest in time to the respective IR images shown in Fig.~\ref{WISESPITZER}.}
\label{HSTPanSTARRS}
\end{figure}

To assess the detectability of coma dust in \textit{Spitzer} data, we measured the FWHM of the PSF of 324P and of isolated stars. We found 1.5 pixels < FWHM < 2 pixels for both 324P and the stars, indicating that the dust of the coma (if present) is unresolved. 
The \textit{Spitzer} images (Fig.~\ref{WISESPITZER}) are characterized by a crowded field of stars. For this reason we refrained from stacking the images and instead performed circular aperture photometry on single images with a fixed physical radius of 3000 km at the comet. To minimize the resulting flux uncertainty, we used small aperture sizes and selected images in which 324P did not overlap neighbouring stars. 

To constrain the dust geometric albedo, we also estimated the brightness of 324P at the visible wavelengths (Fig.~\ref{HSTPanSTARRS}) at the epochs of our two IR observations (Table~\ref{table:1}). As the reference that is closest in time to the 2010 June 9 -- 11 WISE observation, we identified a z-band PS1 image obtained on 2010 June 26. The absolute magnitude derived from this PS1 measurement inside an aperture with radius \SI{3}{\arcsecond} (4847 km) corresponds to an apparent z-band magnitude of (20.4 $\pm$ 1.4) mag, or to a flux of $0.02^{+0.04}_{-0.01}$ mJy, at the time of the WISE observation.

\begin{figure}[t]
\centering
\resizebox{\hsize}{!}{\includegraphics{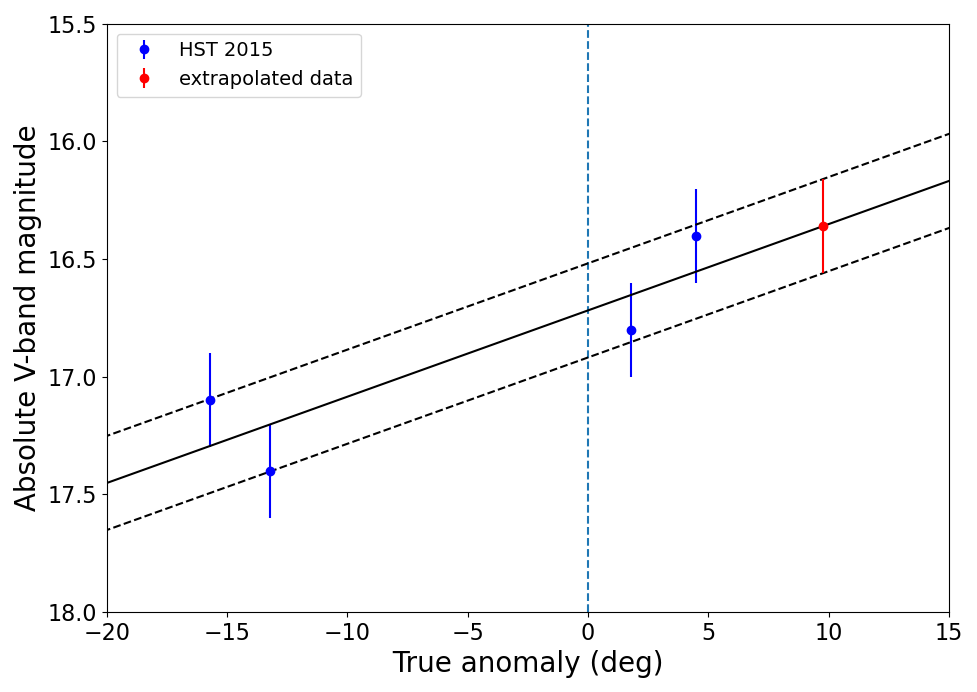}}
\caption{Absolute V-band magnitudes of 324P plotted as a function of the true anomaly and linear fit for extrapolation. The blue dots show HST measurements, and the red dot shows the extrapolated value for 2016 January 8. The central solid line represents the fit to the data, and the dashed lines parallel to the solid line represent the uncertainty of the linear fit.}
\label{extrapolated_data}
\end{figure}

The visible-light reference data for the \textit{Spitzer} epoch that are closest in time are four HST images in F350LP filter. 
To account for the increasing coma brightness between the individual HST observations, 
we extrapolated the absolute magnitudes to 2016 January 8, the epoch of the \textit{Spitzer} measurement (Fig.~\ref{extrapolated_data}) using a linear fit to the absolute magnitudes in Table~\ref{table:1}, 
We calculated the corresponding apparent magnitude at the position of \textit{Spitzer}, finding a value of $(20.9 \pm 1.2)$ mag, and converted it into a flux of $0.02^{+0.03}_{-0.01}$ mJy using
\begin{equation} \label{eq:4}
F=3.622\cdot10^{-5} \cdot 10^{\frac{20-  V  }{2.5}},
\end{equation}
where $F$ is the flux of the source in Jy, $3.622\cdot10^{-5}$ is the flux\footnote{ETC Request IDs: WFC3UVIS.im.1843958, WFC3UVIS.im.1843962.} associated with a Vega magnitude of 20 in Johnson/V filter, and $V$ is the measured magnitude in F350LP filter.

\section{Analysis and results}\label{sec:results}

\subsection{Nucleus magnitude}\label{PhaseFunction}
\begin{figure}[t]
\centering
\resizebox{\hsize}{!}{\includegraphics{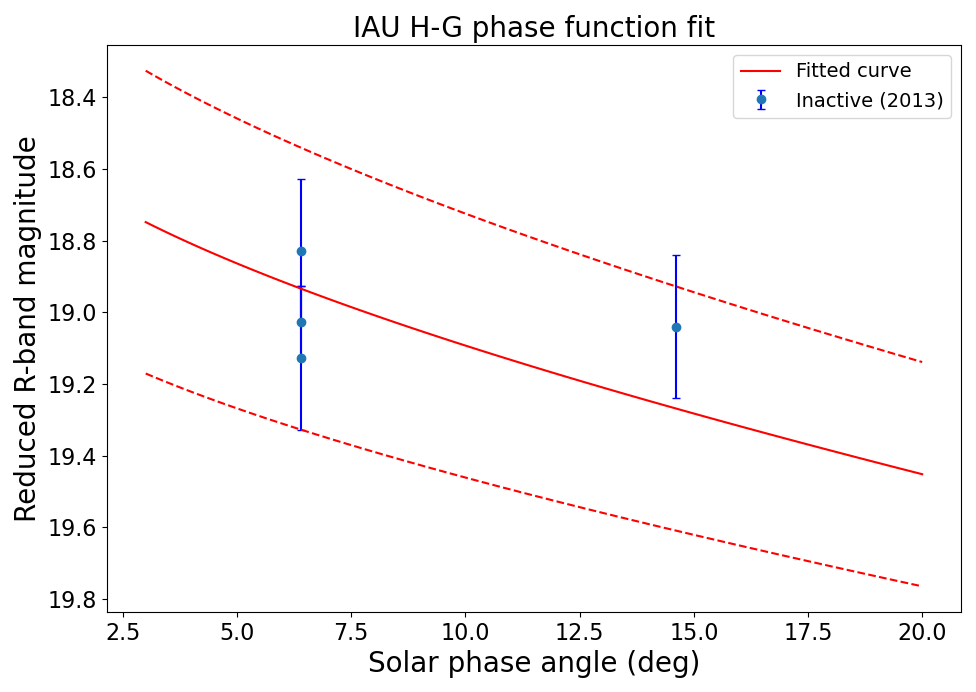}}
\caption{Best-fit IAU phase function (solid red line) for 324P. The dashed red lines represent the uncertainty of the fit. The blue points show 2013 data, when the nucleus was inactive. These were used for the fit. }
\label{IAU}
\end{figure}

To assess the absolute magnitude of the bare nucleus of 324P, we exclusively analyzed data collected in 2013, when no activity was detected. 
We first normalized the measured average apparent R-band magnitudes, $m_{R}$, to unit heliocentric ($r_h$) and geocentric ($\mathit{\Delta}$) distances using 
\begin{equation} \label{eq:norm}
m_{reduced} = m_{R} - 5 \log_{10}(r_h\mathit{\Delta}),
\end{equation}
where $m_{reduced}$ is the reduced R-band magnitude, and $r_h$ and $\mathit{\Delta}$ are measured in AU.
The reduced magnitude is influenced by both the solar phase angle and the rotational phase of the nucleus at the time of observation. The rotation rate of the 324P nucleus is unknown, and our observations are too much scattered in time to constrain it. 
We fit the 2013 reduced magnitudes with an International Astronomical Union (IAU) H-G phase function, and because we lacked appropriate phase angle coverage, we assumed a default C-type value of $G = 0.15 \pm 0.12$ \citep{Bowell1989}, which yielded $H_R = (18.4 \pm 0.5)$ mag and is consistent with the $H_R$ = (18.4 $\pm$ 0.2) mag found by \citet{Hsieh2014}. 
Our assumed uncertainty of $G$ is the standard deviation of many $G$ values found by \citet{Lagerkvist1990} \citep[cf.][]{Polishook2009}. We plot our best-fit phase function in Fig.~\ref{IAU}. 

We calculated the effective nucleus radius using \citep[e.g.][]{Hsieh2023}
\begin{equation} \label{eq:10}
{r_{N}}^2=\frac{(2.24\cdot10^{16})\cdot10^{0.4(m_{\odot, R}-H_{R,inactive})} }{p_R},
\end{equation}
where $p_R = 0.05 \pm 0.02$ is the assumed geometric R-band albedo \citep{albedo, Hsieh2023}, $m_{\odot, R} = -27.15$ is the apparent R-band magnitude of the Sun, and $H_{R,inactive}$ is the absolute R-band magnitude of the inactive nucleus of ($18.4 \pm 0.5$) mag. We found $r_{N}$ = (0.52 $\pm$ 0.16) km. This value agrees with the effective nucleus radius of $r_{N} = 0.59^{+0.18}_{-0.10}$ km found in \citet{Hsieh2023}, who assumed the same V-band albedo of $p_V = 0.05 \pm 0.02$.

\subsection{Visible-light photometry of 324P in active state}\label{absmag}
The average absolute R-band magnitudes, $H_{R}$, were calculated using \citep[e.g.][]{Hsieh2010}
\begin{equation} \label{eq:1}
   H_{R}=m_{reduced} +2.5 \log_{10}[(1-G)\Phi_1(\alpha)+G\Phi_2(\alpha)],
\end{equation}
where $\alpha$ is the solar phase angle and the term $(1-G)\Phi_1(\alpha)+G\Phi_2(\alpha)$ is called the scattering phase function \citep{Bowell1989}. We also used $G = 0.15 \pm 0.12$ (Sect.~\ref{PhaseFunction}) for the dust.
While there are measurements of the dust phase function for comets \citep{Divine1981, lowell_dustphase}, we are not aware of a measured phase function for asteroid dust. Since there are indications that the dust properties for comets and asteroids are different (e.g. from comparing 67P measurements and Ryugu samples), we considered the nucleus phase function to be the most natural proxy for the unknown dust phase function although we are aware that this is most likely an oversimplification.

\begin{figure}[t]
\centering
\resizebox{\hsize}{!}{\includegraphics{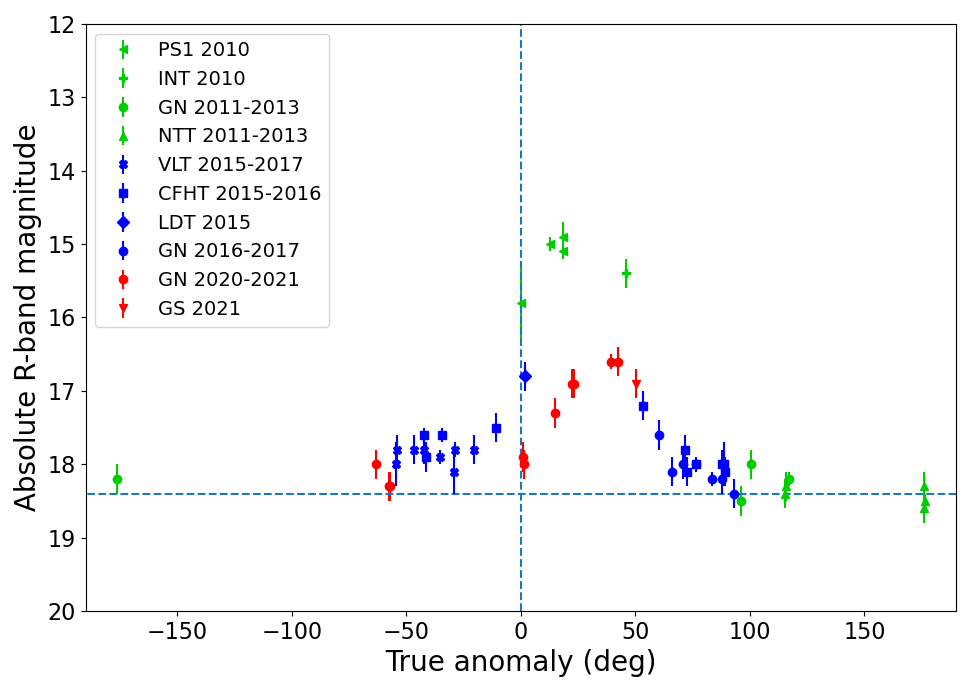}}
\caption{Measured absolute R-band magnitudes of 324P in a 2300 km radius aperture plotted as a function of the true anomaly. The different symbols correspond to the different telescopes and instruments, and colors mark the perihelion passages: green for 2010, blue for 2015, and red for 2021. The vertical dashed line indicates perihelion. The horizontal dashed line corresponds to the absolute magnitude of the inactive nucleus (18.4 mag).}
\label{mag}
\end{figure}

We plot the resulting absolute R-band magnitudes as a function of the true anomaly in Figs.~\ref{mag} and~\ref{colors}, and we list them in Tables~\ref{table:1} and~\ref{5arcsec}.
In Fig.~\ref{colors}, we compare our photometric measurements with literature data. 
The pentagons with black edges were acquired by us with an aperture angular radius of \SI{5}{\arcsecond} instead of 2300 km to enable a direct comparison with \citet{Hsieh2012b} (green crosses in Fig.~\ref{colors} and Table~\ref{5arcsec}). The difference of about one magnitude can be attributed to improvements in all-sky photometric catalogs since the first study. 
Table~\ref{5arcsec} also contains measurements from a \SI{2}{\arcsecond} aperture for five 2015 epochs for comparison with \citet{HsiehSheppard2015}, who used an aperture of the same size, but a different IAU phase function $G$ parameter, $G = 0.17 \pm 0.10$, as in \citet{Hsieh2014}.
In this case, our results agree with theirs.
Fig.~\ref{colors} also includes HST data from \citet{Jewitt2016}, and our results for the same datasets, converted from the V into the R band using $V-R=0.35$, agree well.

\begin{table*}[t]
\centering
\caption{Measurements of the MBC 324P using different apertures.} 
\label{5arcsec}
\begin{threeparttable}
\begin{tabular}{lccrrrrrrr}
\toprule
UT Date & Telescope$^a$ &  N$^b$& $r_h$$^c$ & $\mathit{\Delta}$$^d$ & $\alpha$$^e$ & $\nu$$^f$  & $m_{R}$$^g$ & $H_{R}$$^h$  & $M_{d}$ (x $10^6$)$^i$ \\
\midrule
2010 Aug 16 &PS1 & 4   &  2.631 &  1.795& 15.1& 12.8  &    18.7 $\pm$ 0.1 &  14.5 $\pm$ 0.2& 100 $\pm$ 44 \\ 
2010 Sep 08 &PS1 &  3  &  2.641& 1.739&  	12.0&  18.4& 18.40 $\pm$ 0.05 & 14.4 $\pm$ 0.1& 112 $\pm$ 46 \\ 
2010 Dec 31 &INT &  11  &   2.732& 2.782& 20.5&  45.7& 19.2 $\pm$ 0.1 & 13.8 $\pm$ 0.2& 193 $\pm$ 85  \\ 
2011 Aug 31 &GN &  6  &    3.072& 3.238&	18.2& 95.9& 24.3 $\pm$ 0.2  & 18.4 $\pm$ 0.2 & 0.0 $\pm$ 1.4\\ 
2015 Apr 26   &CFHT& 2 &2.764 &  2.442&  21.1& 307.5& 22.6 $\pm$ 0.2&17.4 $\pm$ 0.2 &4.3 $\pm$ 2.5\\ 
2015 May 16  &CFHT&4&2.740& 2.165&   19.6& 312.0& 22.5 $\pm$ 0.1&17.6 $\pm$ 0.2 &3.1 $\pm$ 2.1\\ 
2015 May 17   &CFHT&2&2.739&  2.152&   19.5& 312.3& 22.4 $\pm$ 0.1&17.6 $\pm$ 0.2 &3.1 $\pm$ 2.1\\ 
2015 May 21  &CFHT&3&2.735&  2.101& 18.0& 313.2& 22.3 $\pm$ 0.1&17.5 $\pm$ 0.2&3.6 $\pm$ 2.3\\ 
2015 May 23  &CFHT&3& 2.732& 2.076&   18.6& 313.6& 22.2 $\pm$ 0.1&17.5 $\pm$ 0.2 &3.6 $\pm$ 2.3\\ 
\bottomrule
\end{tabular}
\begin{tablenotes}\small{
\item$^a$Telescope. $^b$Number of exposures taken. $^c$Heliocentric distance in AU. $^d$Geocentric distance in AU. $^e$Solar phase angle (Sun-Target-Observer) in degrees. $^f$True anomaly in degrees. $^g$Apparent R-band magnitude measured with a \SI{5}{\arcsecond} aperture (2010 and 2011 data) and a \SI{2}{\arcsecond} aperture (2015 data). $^h$Absolute R-band magnitude assuming IAU H-G phase function where $G = 0.15 \pm 0.12$. $^i$Estimated dust mass in kg assuming $\rho_{d}$ $\sim$ 2500 kg m$^{-3}$.\\
Notes: For comparison with published values (see text), the measurements listed here were obtained with a \SI{5}{\arcsecond} aperture for data in 2010 and 2011 and with a \SI{2}{\arcsecond} aperture for data in 2015. Data presented in this table represent the results of our own measurements. 2010 and 2011 datasets were previously published in \citet{Hsieh2012b}, and 2015 datasets in \citet{HsiehSheppard2015}.
}
\end{tablenotes}
\end{threeparttable}
\end{table*}

\subsection{Dependence of the aperture size on absolute dust magnitudes and $Af\rho$}
\label{subsec:afrho}
\begin{figure}[t]
\centering
\resizebox{\hsize}{!}{\includegraphics{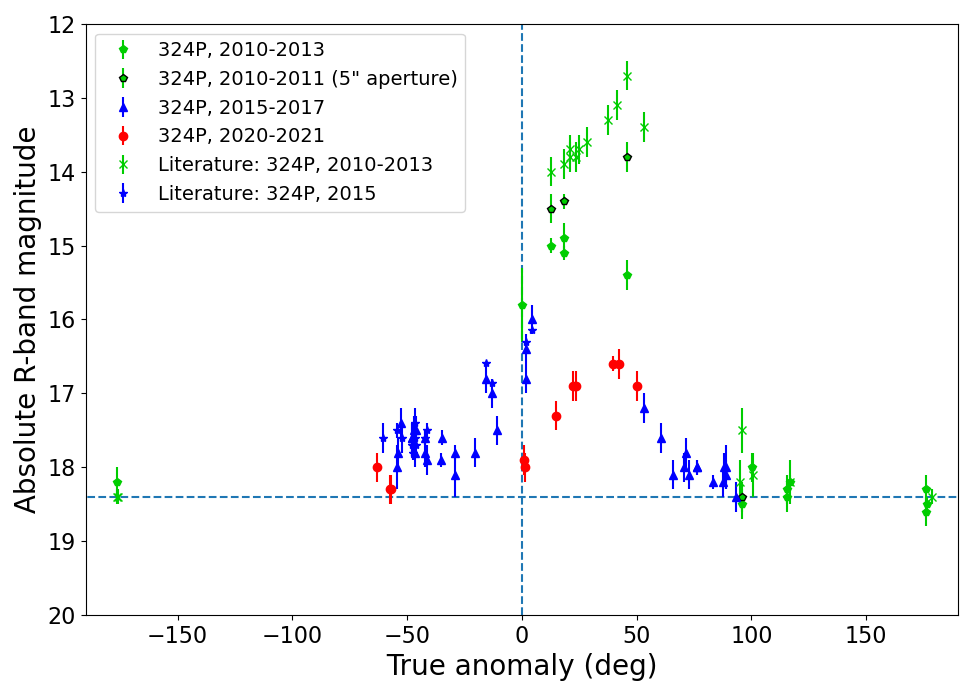}}
\caption{Same as Fig.~\ref{mag}, but augmented by values from the literature (2010-2013: \citet{Hsieh2012b, Hsieh2014}, 2015: \citet{HsiehSheppard2015, Jewitt2016}). Symbols now distinguish between our analysis of archival data and data from the literature (see legend). Pentagons with black edges were measured by us using a \SI{5}{\arcsecond} aperture for comparison with \citet{Hsieh2012b} (green crosses).}
\label{colors}
\end{figure}

\begin{figure*}[t]
\centering
\resizebox{\hsize}{!}{\includegraphics{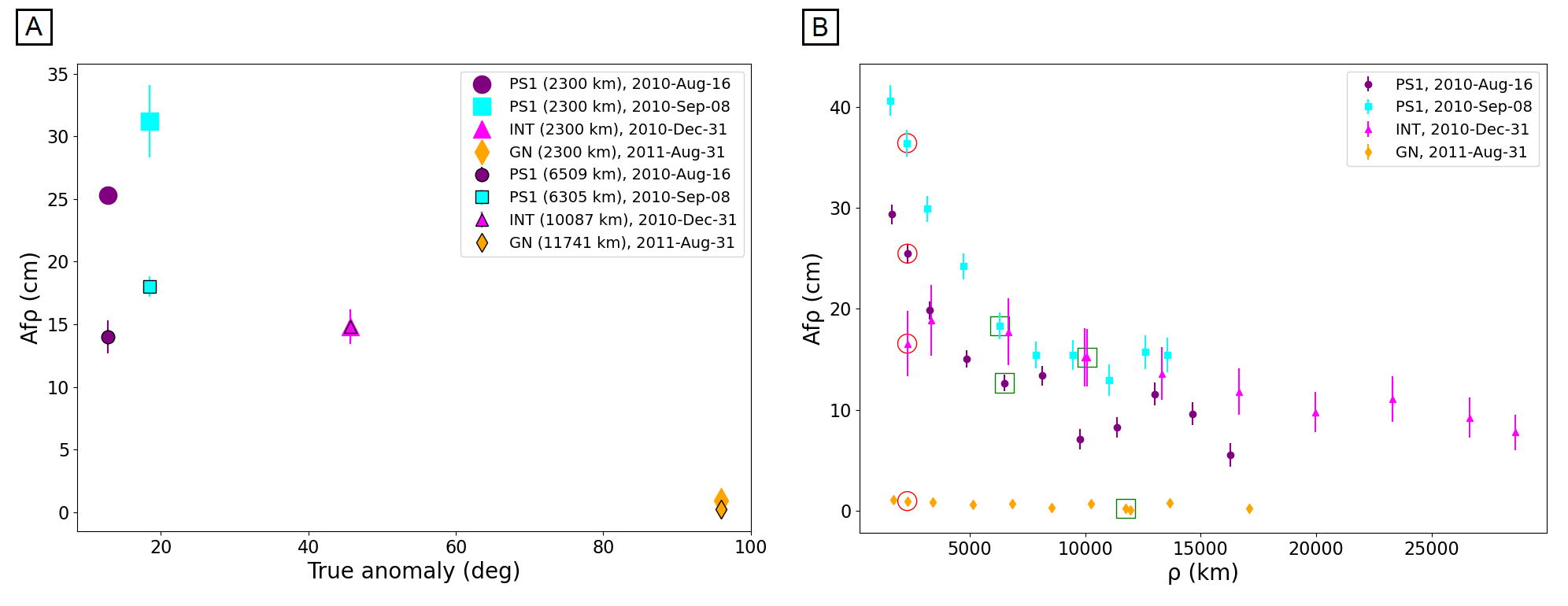}}
\caption{$Af\rho$ parameter. Panel A: $Af\rho$ vs. true anomaly for 324P, evaluated using circular apertures with a physical radius of 2300 km (symbols without borders) and circular apertures with angular radius \SI{5}{\arcsecond} (symbols with black borders). Panel B: $Af\rho$ vs. aperture radius $\rho$. The data enclosed by red circles correspond to the border-less symbols of panel A ($\rho$ = 2300 km), and data enclosed by green squares correspond to the black-bordered data points of panel A (angular radius of \SI{5}{\arcsecond}).}
\label{afrho}
\end{figure*}

We now compare the time evolution of the magnitudes measured in  differently sized apertures (Fig.~\ref{colors}). At $\nu$ = 45.7°, the brightness in the larger aperture (5\arcsec, black pentagons) has increased compared to earlier measurements, and the brightness in the smaller aperture (2300 km, green pentagons) has decreased. One possible cause of this discrepancy could be the changing geocentric distance, which affects the physical volume enclosed in an aperture with a fixed angular size. At $\mathit{\Delta}$ = 1.74 AU ($\nu = 18^{\circ}$), less volume is enclosed in the 5\arcsec\ aperture than at 2.78 AU ($\nu = 45^{\circ}$), so that the absolute magnitude of dust in an aperture with a fixed angular size is expected to decrease with increasing distance to the observer. 
To better understand the dependence of the enclosed dust cross section on the aperture size and time, we calculated the $Af\rho$ parameter (Table~\ref{table:Afrho} and Fig.~\ref{afrho}), which for a coma in steady state that is spherically symmetric is expected to be independent of the aperture size. 
$Af\rho$ is the product of the albedo $A$, defined as the total light reflected by the cometary grains over the total light received, the filling factor $f$ of the grains in the field of view, and $\rho$, the physical radius of the field of view \citep{Fink2012}. We used \citep{A'Hearn1984}
\begin{equation} \label{eq:11}
Af\rho = \frac{(2 r_h \mathit{\Delta})^2}{\rho} 10^{0.4(m_{\odot, R}-m_{R})},
\end{equation}
where $r_h$ is the heliocentric distance in AU, $\mathit{\Delta}$ is the geocentric distance in cm, $\rho$ is the physical radius in cm at the distance of the MBC, $m_{\odot, R}=-27.15$ is the apparent R-band magnitude of the Sun, and $m_{R}$ is the measured apparent R-band magnitude of the MBC.

Plotting $Af\rho$ versus true anomaly (Fig.~\ref{afrho}, Panel A), we find that at $\nu$ = 45.7°, $Af\rho$ is similar from both apertures (\SI{5}{\arcsecond}, and 2300 km). 
This behavior is consistent with a steady-state coma, and the difference of $>$ 1 mag in Fig.~\ref{colors} results from the spatially extended nature of the dust coma, where larger apertures enclose more dust.

\begin{table}[t]
\caption{$Af\rho$ parameter.}
\label{table:Afrho}
\begin{threeparttable}
\setlength{\tabcolsep}{2pt}
\begin{tabular}{lcrr}
\toprule
UT Date&Telescope$^a$&$Af\rho$ (for 2300 km)$^b$ &$Af\rho$ (for $\SI{5}{\arcsecond})$$^c$\\
\midrule
2010 Aug 16& PS1&  25.3 $\pm$ 0.5& 14.0 $\pm$ 1.3\\
2010 Sep 08& PS1&  31.2 $\pm$ 2.9& 18.0 $\pm$ 0.8\\
2010 Dec 31 & INT&  14.8 $\pm$ 1.4& 14.8 $\pm$ 1.4\\
2011 Aug 31& GN&  0.9 $\pm$ 0.1& 0.20 $\pm$ 0.04\\
\bottomrule
\end{tabular}
\begin{tablenotes}\small{
\item$^a$Telescope. $^b$$Af\rho$ in cm measured for a 2300 km physical radius. $^c$$Af\rho$ in cm measured for a $\SI{5}{\arcsecond}$ angular size. \\
Note: $Af\rho$ parameter for the epochs described by the first four lines of Table~\ref{5arcsec} from two different aperture sizes.
}
\end{tablenotes}
\end{threeparttable}
\end{table}

At true anomalies 12.8° and 18.4°, $Af\rho$ increases with decreasing aperture size, indicating that the dust density decreases more steeply with increasing nucleus distance than in steady state. During this early phase of the activity, the dust production rate might still have been increasing with time, such that freshly emitted dust located close to the nucleus would be relatively more abundant than dust emitted at an earlier stage that had traveled farther at the time of observation.
Hence, $Af\rho$ derived from the smaller aperture would be higher.

However, the $Af\rho$ parameter has some limitations: It assumes that there is no dust production or destruction (e.g., by fragmentation or sublimation of embedded ice) after the dust particles leave the nucleus, it assumes that the dust has a constant outflow velocity, and it fails at the turnaround distance \citep{Fink2012}. 
In panel B of Fig.~\ref{afrho}, we plot $Af\rho$ versus the aperture radius $\rho$ at four epochs.
The plot again shows a decrease in $Af\rho$ with aperture size when 324P was freshly active in August and September 2010, a more shallow profile at the end of December 2010, and a flat, almost zero profile in August. 

The flattening of the $Af\rho$ profile at sufficiently large aperture radii during the first two epochs may suggest that during an even earlier epoch, the coma was closer to a steady-state regime than at the times of observation. This is consistent with Fig.~\ref{mag}, which shows a flat profile of absolute coma magnitudes up to a true anomaly of about 20° and then a steep increase.

In panel B of Fig.~\ref{afrho}, at large $\rho$, the $Af\rho$ parameter exhibits strong fluctuations that are likely due to an increasingly variable background flux as the aperture size increases. However, these fluctuations occur only beyond aperture radii of \SI{5}{\arcsecond} and are therefore not expected to significantly affect our interpretation of panel A of Fig.~\ref{afrho}.

\subsection{Dust-mass estimates and onset of activity}
\begin{figure}[t]
\centering
\resizebox{\hsize}{!}{\includegraphics{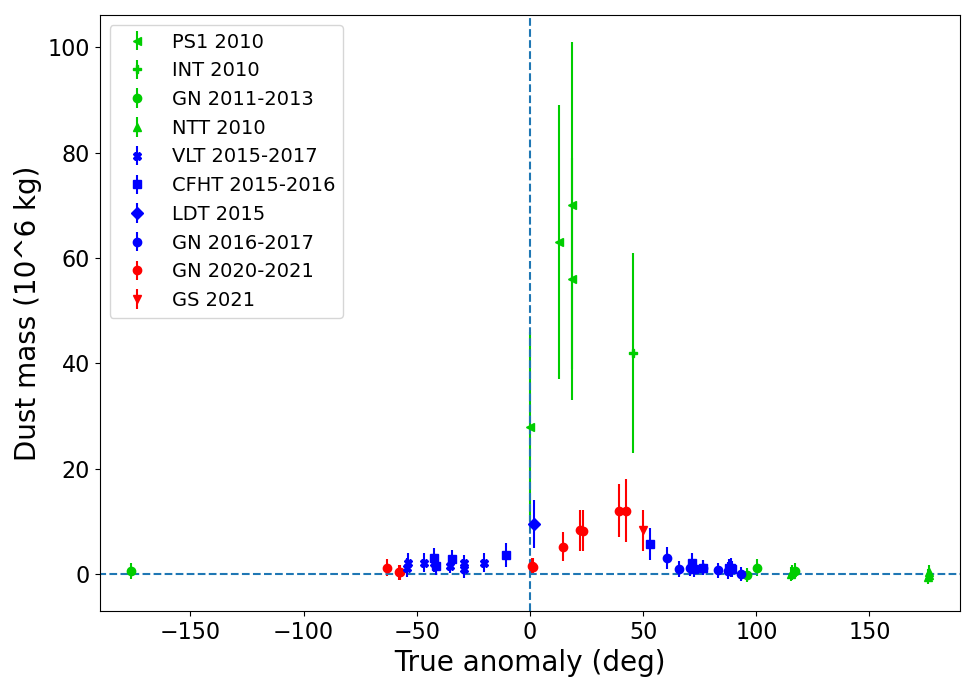}}
\caption{Estimated dust masses for 324P plotted as a function of the true anomaly. The different symbols correspond to the different telescopes and instruments used, and the different colors correspond to the three different perihelion passages: green for 2010, blue for 2015, and red for 2021. The horizontal dashed line corresponds to a zero dust mass (the MBC was not active).}
\label{dust}
\end{figure}

We estimated the dust mass inside the aperture using \citep[e.g.][]{HsiehSheppard2015}
\begin{equation} \label{eq:2}
M_d=\frac{4}{3} \pi r_N^2 a \rho_{d}	    \left(  \frac{1-10^{  0.4(H_{R}-H_{R,inactive})  }  }{10^{  0.4(H_{R}-H_{R,inactive}) }}\right),
\end{equation}
where $r_N = 0.52$ km is the estimated effective nucleus radius for 324P (Sect.~\ref{PhaseFunction}), $a = 1$ mm is the assumed effective mean dust grain radius, $\rho_{d}$ $\sim$ 2500 kg m$^{-3}$ is the assumed dust grain density (C-type objects), and $H_{R,inactive} = (18.4 \pm 0.5)$ mag is the absolute magnitude of the inactive nucleus of 324P in the R band (Sect.~\ref{PhaseFunction}). 
We plot our estimated dust masses as a function of the true anomaly in Fig.~\ref{dust}. 
The graphic shows that the MBC starts to be active near perihelion, and this suggests that the activity is due to sublimation. 
In Table~\ref{table:1} we list the results of the photometric analysis. Three dust-mass values are negative but still consistent with zero, given the uncertainties.

Using a linear fit to the absolute magnitude as a function of time, \citet{HsiehSheppard2015} found the average net dust production rate ($\dot{M_d}$) in 2015 for 324P to be $\lesssim$ 0.1 kg s$^{-1}$ (-60.4° < $\nu$ < -41.4°), which is lower by almost 2 orders of magnitude than the net dust production rate value of $\sim$ 30 kg s$^{-1}$ (12.9° < $\nu$ < 45.9°) measured in 2010 by \citet{Hsieh2014}. These dust production rates from the literature were measured at different points in the orbit and were measured with rectangular apertures, aiming to capture the total visible flux (including coma and tail).

\begin{figure}[t]
\centering
\resizebox{\hsize}{!}{\includegraphics{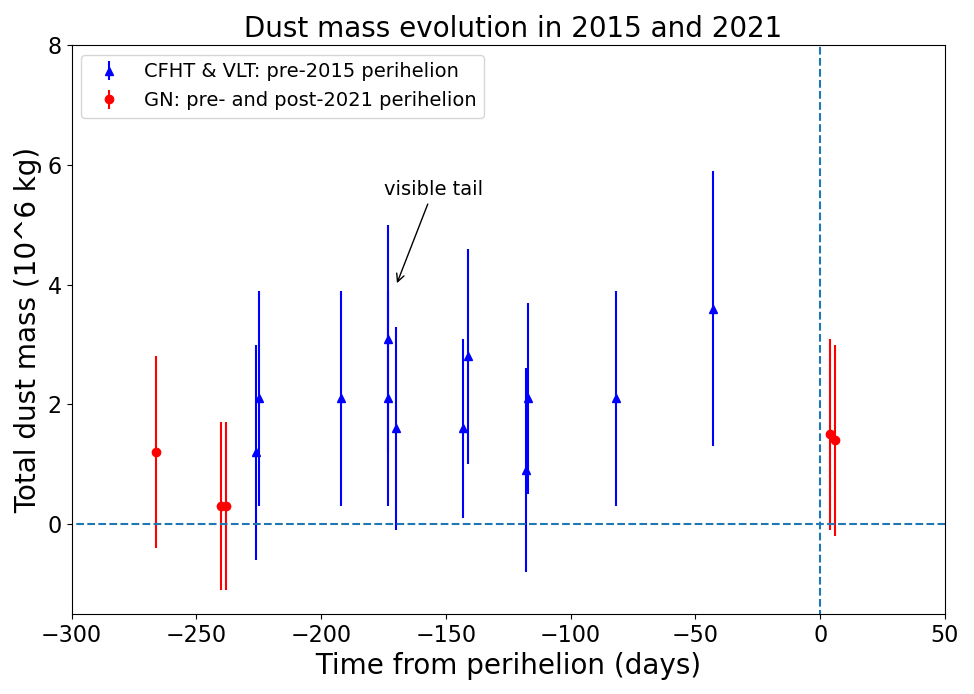}}
\caption{Estimated dust masses around 324P plotted vs. time from perihelion.
Pre-perihelion data in 2015 (blue) and pre- and post-perihelion data in 2021 (red). The epoch when the tail starts to become visible is marked in the plot.} 
\label{FigX}
\end{figure}

In our study, however, this linear fitting approach was not applicable because we observed nearly constant values in our dust measurements (Fig.~\ref{FigX}). 
When we assume that dust particles are ejected from 324P with a speed comparable to the nucleus gravitational escape speed of $v_{escape} = 0.61$ m s$^{-1}$ (derived with the estimated nucleus size and assumed bulk density), they leave the 2300 km aperture after 43 days.

The data points displayed in Fig.~\ref{FigX} were collected over a period of several hundred days, suggesting that the earliest ejected dust had already left the measurement aperture at most epochs. Consequently, the nearly constant dust mass suggests that the rate of dust production from 324P may be roughly equal to the rate of dust loss from the aperture.

For the 2015 perihelion, we infer that the onset of activity took place earlier than 173 days before perihelion, when the image in Fig.~\ref{mosaic} shows a visible tail on 2015 June 10 (marked in Fig~\ref{FigX}).
From the same criterion, we cannot infer anything about the onset time for the 2021 perihelion passage because no tail is visible in Fig.~\ref{mosaic} in the 2020 data.

In Fig.~\ref{colors}, near $\nu$ = 120°, the absolute magnitude stabilizes around 18.4 mag in the R band, consistent with our derived magnitude of the inactive nucleus (Sect.~\ref{PhaseFunction}). This indicates that the dust production has ceased and that dust has left the immediate environment of the nucleus due to solar radiation pressure (in 2011).

\begin{figure}[t]
\centering
\resizebox{\hsize}{!}{\includegraphics{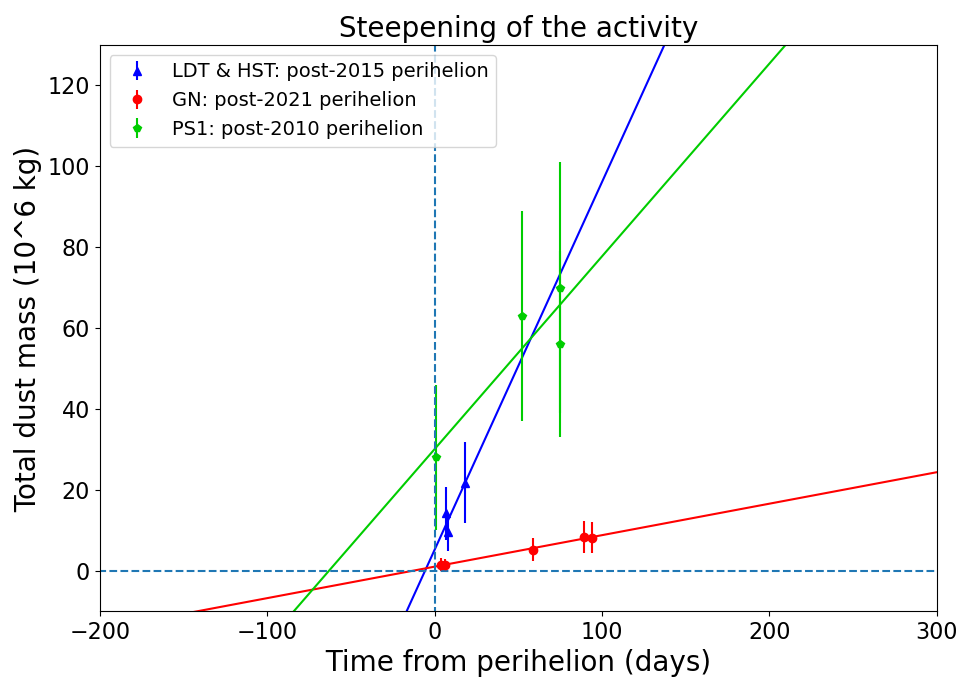}}
\caption{Estimated dust masses around 324P plotted vs. time from perihelion. The dashed lines show linear fits to post-perihelion data from 2010 (green), 2015 (blue) and 2021 (red), which were used to calculate the time at which the activity increased.}
\label{cresce}
\end{figure}

Fig.~\ref{colors} shows a steep increase in brightness at about and immediately after perihelion. To find the onset time of this steepening, we fit post-perihelion dust masses with a linear function. We determined the onset times of the profile steepening to be (33 - 81) days before perihelion in 2010 with a mass-loss rate of $(5.5 \pm 1.4)$ kg s$^{-1}$ (0.0° < $\nu$ < 18.4°), 6 days before perihelion in 2015 with a mass-loss rate of $(10 \pm 4)$ kg s$^{-1}$ (1.8° < $\nu$ < 4.5°), and between 23 days before perihelion and 8 days after perihelion in 2021 with a mass-loss rate of $(0.9 \pm 0.3)$ kg s$^{-1}$ (1.0° < $\nu$ < 23.4°) (Fig.~\ref{cresce}). This steep rise in activity is likely reflected in the $Af\rho$-profiles as discussed in Section~\ref{subsec:afrho}. It may be caused by the thermal wave reaching a subsurface layer with elevated ice content, or by seasonal exposure of an ice reservoir that is not reached by sunlight before perihelion.

\citet{Hui2017} studied data from 2010 to 2015 that indicated that 324P exhibited nongravitational accelerations caused by recoil forces due to anisotropic mass loss, with dynamically inferred mass-loss rates of $(36 \pm 3)$ kg s$^{-1}$. This value agrees with the findings of \citet{Hsieh2014} in 2010. However, our analysis of the 2010 data reveals mass-loss rates that differ by nearly an order of magnitude, which we discussed in Sect.~\ref{absmag} as possibly resulting from advances in all-sky photometric catalogs. Additionally, our estimated mass-loss rate agrees with that reported by \citet{Moreno2011}, who found a value of 3 – 4 kg s$^{-1}$ between 2010 October and 2011 January (26.5° < $\nu$ < 47.8°).

Figs.~\ref{colors} and~\ref{dust} visually confirm that the activity during the 2015 perihelion passage significantly decreased compared to the previous perihelion passage in 2010. The activity in 2021 also decreased compared to 2015. 

A possible interpretation of the decrease in activity from one perihelion passage to the next may be given by the buildup of a dry mantle. Globally, ice-rich main-belt asteroids are thought to have a rubble mantle-type crust. Following the trigger event, the freshly exposed ice in an MBC sublimates and causes the observable activity. Dust particles that are too large to be ejected by the gas drag against gravity would accumulate on the surface. Gradually, this layer of dust and pebbles again becomes thick enough to act as a protective layer and buries the ice in the depths \citep{Thiel1989, Jewitt1992, Jewitt1996}. 
In this scenario, the activity in MBCs should initially decrease rapidly after the trigger event. Subsequently, as a rubble mantle of sufficient thickness forms, the activity should decrease more slowly \citep{Jewitt1996, Hsieh2015, Hsieh2018a}. This might be a possible explanation for the rapid decrease in the activity of 324P in 2015 compared to that in 2010 and the following slow decrease of its activity in 2021. Data from upcoming perihelia are required to corroborate this hypothesis, however.

\subsection{Analysis of infrared data}\label{PhotometricAnalysisIR}

\subsubsection*{Thermal and scattered components} \label{SEDsection}
Infrared images provide information about the thermal emission from the grains \citep{Sarmecanic1997}, while visual images provide information about the reflectivity of the grains \citep{Kolokolova2004}. By combining these pieces of information, we calculated the geometric albedo of the cometary grains, which is defined as the ratio of the amount of light that is scattered by the grains to the amount of light that would be scattered to zero phase angle by a perfectly reflective, isotropic surface, known as a Lambertian surface \citep{Hanner1981}.

The thermal component is described by \citep{JewittMeech1988}
\begin{equation} \label{eq:7}
   F_{IR}=   \frac{ \varepsilon S B_{\nu }(T)}{\mathit{\Delta}^2},
\end{equation}
where $F_{IR}$ is the thermal flux density from the aperture in W m$^{-2}$ $\mathrm{Hz}^{-1}$, $\varepsilon$ is the emissivity, assumed here to equal unity, 
$S$ is the cross section of the cometary grains inside the aperture in m${^2}$, $B_{\nu }(T)$ is the Planck function evaluated at the temperature $T$ in W m$^{-2}$ $\mathrm{Hz}^{-1}$ sr$^{-1}$, and $\mathit{\Delta}$ is the geocentric distance in meters.

The scattered component is given by \citep{Russell1916}
\begin{equation} \label{eq:8}
F_{\rm vis} = F_\odot \frac{p j(\alpha) S}{(r_{h}/1 \mathrm{AU})^2\pi \mathit{\Delta}^2},
\end{equation}
where $F_{\odot}$ is the solar spectrum at 1 AU and at the wavelength used for the measurement of the flux from the MBC, $F_{\rm vis}$, and both are in the same units.
The quantity $p$ is the geometric albedo of the dust particles, $j(\alpha)$ is the scattering phase function, and $r_{h}$ and $\mathit{\Delta}$ are the heliocentric and geocentric distances in meters, respectively.

\begin{figure}[t]
\centering
\resizebox{\hsize}{!}{\includegraphics{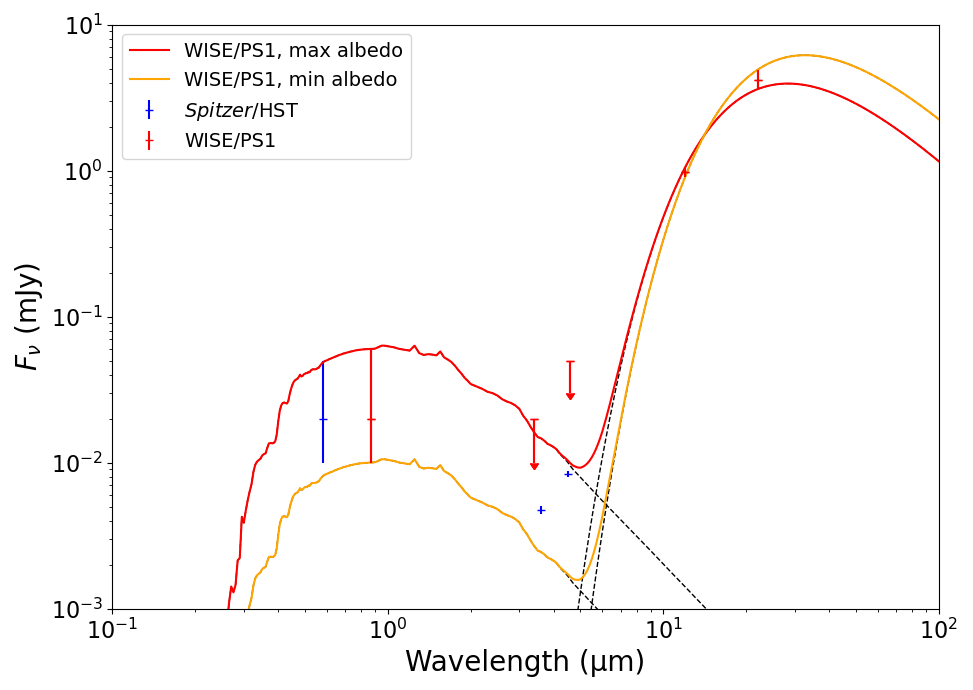}}
\caption{SEDs corresponding to the maximum (red) and minimum (yellow) derived geometric albedo values from the WISE/PS1 combination. The \textit{Spitzer}/HST data are also plotted (blue). 
Even though taken during different perihelion passages and from different observer distances, we plot the measurements and resulting fits in a single graph. The dashed lines indicate the pure blackbody and solar spectra composing the overall SED.}
\label{fig:sed}
\end{figure}

The spectral energy distribution (\citealp[SED;][]{Williams1997, Kolokolova2004, Yang2009}) is composed of the superposition of scattered sunlight and thermal emission by the dust (Fig.~\ref{fig:sed}). 
From the shape of the thermal SED, we first derived the dust temperature, then the product of dust cross section and emissivity, and the scattered component finally gives the product of geometric albedo and phase function divided by the emissivity at the time of observation of the cometary grains.

We derived the dust color temperature from equating the flux ratio in the WISE W4/W3 bands to the ratio of the Planck function at these wavelengths, $B_{\nu }$(22 µm, $T$) / $B_{\nu }$(12 µm, $T$), with the temperature $T_{col}$ as a free parameter, and obtained $T_{col} = 167^{+13}_{-11}$ K. This temperature range includes the equilibrium temperature of a fast-rotating sphere with an isothermal surface, $T_{eq}$=172 K, at the heliocentric distance during the WISE observation $r_{h} = 2.623$ AU, with

\begin{equation} \label{eq:8.5}
 T_{eq}(r_h, A_B, \varepsilon)=278.8 \left(\frac{1-A_B}{\varepsilon}\right)^{\frac{1}{4}} \frac{1}{\sqrt{r_h}},
\end{equation}
where $A_B$ is the Bond albedo, assumed to be 0, $\varepsilon$ is the emissivity, assumed to be 1, and $r_h$ is the heliocentric distance in AU. 
We conclude that there is no strong indication that dust superheats at mid-infrared (mid-IR) wavelengths, which is consistent with compact or large particles, or both \citep{GehrzNey1992}, but that color temperatures up to 8\% above $T_{eq}$ are consistent with the WISE measurement.

To fit the combined SED of scattered light and thermal emission to the PS1 and WISE data, we scaled the Planck function, $B_\nu$, with a freely variable factor $f_{IR}$ and the solar spectrum\footnote{\url{https://www.nrel.gov/grid/solar-resource/spectra.html},  $F_{\odot}$, the Thekaekara Spectrum.} with an also freely variable factor $f_{vis}$, such that their sum matches the measured fluxes.

From Eq.~(\ref{eq:7}), the scaling factor (in units of sr) $f_{IR}$ is equal to 
\begin{equation} \label{eq:7scaling}
   f_{IR}(T_{col}) =   \frac{\varepsilon S(T_{col})}{\mathit{\Delta}^2}.
\end{equation}
The $S$ implicitly depends on the color temperature for a given measured flux $F_{IR}$ because of the temperature dependence of $B_\nu(T)$ in Eq.~(\ref{eq:7}).
From Eq.~(\ref{eq:8}), the dimensionless scaling factor for the scattering component is equal to 
\begin{equation} \label{eq:8scaling}
f_{vis} =   \frac{p j(\alpha) S}{(r_{h}/1 \mathrm{AU})^2\pi \mathit{\Delta}^2},
\end{equation}
such that 
\begin{equation}
\frac{p j(\alpha)}{\varepsilon} = \pi (r_h/1 \mathrm{AU})^2 \frac{f_{vis}}{f_{IR}},
\label{eq:pjeps}
\end{equation}
assuming that the amount of dust seen in paired images is the same, despite different PSF and/or aperture sizes and observation dates.
The maximum scaling factor, $f_{IR}^{max}$, is obtained for the minimum possible color temperature, and vice versa. The minimum and maximum values for $f_{vis}$ result from fitting the solar spectrum to the lower and upper bound of the PS1 measurement error bar. Inserting these into Eq.~(\ref{eq:pjeps}), we find $p j(\alpha)$ = 0.03$^{+0.12}_{-0.02}$. This corresponds to 3\% $<p<$ 45\% for a C-type ($G = 0.15 \pm 0.12$) phase function, and 2\% $<p<$ 40\% for an S-type ($G = 0.25 \pm 0.12$). Since this range includes all typical values measured for both C- and S-types, we cannot meaningfully constrain the geometric albedo or the spectral type of the dust.
These large errors in the albedo calculation are due to the wide range of temperatures consistent with the mid-IR data and to large uncertainties in the photometry at optical wavelengths. 

Fig.~\ref{fig:sed} shows the minimum and maximum geometric albedo SEDs derived from the WISE/PS1 dataset together with these data, and in addition, the \textit{Spitzer} and the extrapolated HST data (flux values in Tables~\ref{datadots1} and~\ref{datadots2}). The latter were obtained during a different perihelion passage, when the overall dust brightness was lower than in 2010 (see~Fig.~\ref{colors}), and from a different observer distance, but at very similar heliocentric distance, phase angle, and true anomaly. We show all six data points in a single graph because (1) the similar visible-light fluxes from PS1 and HST indicate that the overall amount of dust, scaled with $\mathit{\Delta}^2$, during both measurements was similar, (2) the similar heliocentric distances indicate that the color temperature range during both measurements was likely similar, and (3) the comparison with the WISE SEDs can give us some directions for the interpretation of the \textit{Spitzer} data.

\begin{table}[t]
\begin{threeparttable}
\caption{Telescope and band of observation for the \textit{Spitzer}/HST combination.}
\label{datadots1}
\setlength{\tabcolsep}{2.3pt}
\begin{tabular}{lccc}
\toprule
Telescope: &HST &\textit{Spitzer} &\textit{Spitzer}  \\
\midrule
Wavelength: &0.58 µm& 3.6 µm& 4.5 µm\\
\midrule
Flux (mJy):&$0.02^{+0.03}_{-0.01}$ &0.0047 $\pm$ 0.0003&0.0084 $\pm$ 0.0004  \\
\bottomrule
\end{tabular}
\begin{tablenotes}\small{
    \item Note: Combination for 2016 January 8 using 3000 km apertures throughout.}
\end{tablenotes}
\end{threeparttable}
\end{table}

The plot shows that the longer of the two IRAC wavelengths lies in the overlapping zone where both scattered light and thermal emission can contribute significantly to the overall flux, while the flux measured at the shorter wavelength is most certainly dominated by scattered light, such that an independent measurement of the color temperature from the IRAC data is not possible. 
The dust brightness in the 4.5 µm IRAC measurement is also inconsistent with our simple model of representing the thermal emission by a single blackbody function. A possible cause for the elevated flux at 4.5 µm could be superheating because small or porous dust particles may not be able to thermally radiate efficiently at shorter wavelengths. Excess brightness in the IRAC 4.5 µm band has also been interpreted as an indicator for the presence of CO or CO$_2$ vapor \citep{Ootsubo2011}, which we cannot exclude. However, our measured ratio of $F$(4.5 µm / 3.6 µm) = 1.8 at $r_h$ = 2.625 AU is near the lower limit measured in 23 comets by \citet{reach-kelley2013}, where a low value of this ratio correlates with the absence of PSF broadening attributable to a spatially extended gas coma. We found no indications of an extended coma either, which again does not support the interpretation of our 4.5 µm excess flux as due to CO or CO$_2$ emission.
Depletion of CO$_{2}$ relative to water was also observed in the MBC 238P by JWST \citep{Kelley2023}.

\begin{table}[t]
\begin{threeparttable}
\caption{Telescope and band of observation for the WISE/PS1 combination.}
\label{datadots2}
\setlength{\tabcolsep}{2.3pt} 
\begin{tabular}{lccccc}
\toprule
Telescope:  &PS1    &WISE&WISE  &WISE &WISE \\
\midrule
Wavelength:& 0.87 µm   & 3.4 µm& 4.6 µm & 12 µm& 22 µm\\
\midrule
Flux (mJy):&  $0.02^{+0.04}_{-0.01}$ & $< 0.02$    &   $< 0.05$    &    $1.0 \pm 0.1$    &   $4.2^{+0.7}_{-0.6}$    \\
\bottomrule
\end{tabular}
\begin{tablenotes}\small{
    \item Note: Combination for 2010 June 10 using apertures of 19230 km (W3), 38460 km (W4) and 4847 km for PS1.}
\end{tablenotes}
\end{threeparttable}
\end{table}

\section{Summary}
\label{sec:summary}
We analyzed archival data of the MBC 324P in the period 2010-2021. We list our main results below.
\begin{itemize}
\item Using only photometry measured when the MBC was observed to be inactive, we find the best-fit IAU phase function parameter $H_R = (18.4 \pm 0.5)$ mag using $G = 0.15 \pm 0.12$.
\item We calculated the effective nucleus radius, finding a value of $r_{N} = (0.52 \pm 0.16)$ km (assuming geometric albedo $p_R = 0.05 \pm 0.02$).  
\item We observed a decrease in its activity during the 2015 and 2021 perihelion passages compared to the previous passage in 2010. This decrease in the activity strength might be due to mantling and/or volatile depletion.
\item The study of the $Af\rho$ profile and the absolute magnitudes of the dust coma across different true anomalies uncovered evidence that the coma of 324P transitioned near perihelion from a pre-perihelion steady state to a higher-activity post-perihelion steady state. This might indicate that a thermal wave arrived at more ice-rich layers, or it might indicate seasonal effects.
\item The study of the IR data yielded a dust geometric albedo in the range of (2 - 45)\%, with unclear spectral type, but it suggested excess radiation at 4.5 µm in comparison to a single-temperature blackbody spectrum.
\end{itemize}

Further observations are necessary to monitor the future activity to constrain the possible causes for the decrease in the activity of 324P over time. 
On 2026 October 14, 324P will be at perihelion, and this may provide opportunities to gather further observations, data, and results. Furthermore, a data analysis of 324P during its inactive phases might enhance the accuracy of the IAU phase-function fitting. This might lead to an improved precision in determining absolute magnitudes and dust masses.\\

\begin{acknowledgements}
We thank the anonymous referee for their constructive input. We thank David Jewitt for his feedback on the manuscript, and the WISE Help Desk and the Pan-STARRS1 Help Desk for the support and information exchanged via mail.
MM, YK and JA were funded by VolkswagenStiftung.
The Pan-STARRS1 Surveys (PS1) and the PS1 public science archive have been made possible through contributions by the Institute for Astronomy, the University of Hawaii, the Pan-STARRS Project Office, the Max-Planck Society and its participating institutes, the Max Planck Institute for Astronomy, Heidelberg and the Max Planck Institute for Extraterrestrial Physics, Garching, The Johns Hopkins University, Durham University, the University of Edinburgh, the Queen's University Belfast, the Harvard-Smithsonian Center for Astrophysics, the Las Cumbres Observatory Global Telescope Network Incorporated, the National Central University of Taiwan, the Space Telescope Science Institute, the National Aeronautics and Space Administration under Grant No. NNX08AR22G issued through the Planetary Science Division of the NASA Science Mission Directorate, the National Science Foundation Grant No. AST–1238877, the University of Maryland, Eotvos Lorand University (ELTE), the Los Alamos National Laboratory, and the Gordon and Betty Moore Foundation. PanSTARRS image identifiers: skycell.1760.023, skycell.1926.046, skycell.1926.047, skycell.1925.047. 
Based on observations made with the Isaac Newton Telescope under Director’s Discretionary Time of Spain’s Instituto de Astrofísica de Canarias. INT program: I/2010B/P14 (PI: Hsieh). 
Based on observations obtained at the international Gemini Observatory, a program of NSF’s NOIRLab (acquired through the Gemini Observatory Archive at NSF’s NOIRLab and processed using DRAGONS (Data Reduction for Astronomy from Gemini Observatory North and South)), which is managed by the Association of Universities for Research in Astronomy (AURA) under a cooperative agreement with the National Science Foundation on behalf of the Gemini Observatory partnership: the National Science Foundation (United States), National Research Council (Canada), Agencia Nacional de Investigaci\'{o}n y Desarrollo (Chile), Ministerio de Ciencia, Tecnolog\'{i}a e Innovaci\'{o}n (Argentina), Minist\'{e}rio da Ci\^{e}ncia, Tecnologia, Inova\c{c}\~{o}es e Comunica\c{c}\~{o}es (Brazil), and Korea Astronomy and Space Science Institute (Republic of Korea). Program IDs: GN-2011B-Q-17, GN-2013A-Q-102, GN-2016B-LP-11, GN-2020B-LP-104, GN-2021A-LP-104, GN-2021B-LP-104 and GS-2021B-LP-104 (PI: Hsieh). 
Based on observations made with the New Technology Telescope (NTT) at the ESO La Silla Observatory under the ESO program identifications 184.C-1143(E) and 184.C-1143(G) (PI: Hainaut). 
Based on observations collected at the European Organisation for Astronomical Research in the Southern Hemisphere (processed using EsoRex (ESO Recipe Execution Tool)) under ESO program identification 095.C-0932(A) (PI: Hsieh). 
Based on observations obtained with MegaPrime/MegaCam, a joint project of CFHT and CEA/DAPNIA, at the Canada-France-Hawaii Telescope (CFHT) which is operated by the National Research Council (NRC) of Canada, the Institut National des Science de l'Univers of the Centre National de la Recherche Scientifique (CNRS) of France, and the University of Hawaii. The observations at the Canada-France-Hawaii Telescope were performed with care and respect from the summit of Maunakea which is a significant cultural and historic site. Proposal IDs: 15AT05, 15BT12, 16AT06 and 16BT05 (PI: Hsieh). 
These results made use of the Lowell Discovery Telescope (LDT) at Lowell Observatory.  Lowell is a private, non-profit institution dedicated to astrophysical research and public appreciation of astronomy and operates the LDT in partnership with Boston University, the University of Maryland, the University of Toledo, Northern Arizona University and Yale University. The Large Monolithic Imager was built by Lowell Observatory using funds provided by the National Science Foundation (AST-1005313). Program ID: L02 (PI: Knight). We acknowledge Matthew M. Knight for providing the LDT images used in this study. 
This research is based on observations made with the NASA/ESA \textit{Hubble} Space Telescope obtained from the Space Telescope Science Institute, which is operated by the Association of Universities for Research in Astronomy, Inc., under NASA contract NAS 5–26555. These observations are associated with programs 14263 and 14458 (PI: Jewitt). 
This publication makes use of data products from the Wide-field Infrared Survey Explorer, which is a joint project of the University of California, Los Angeles, and the Jet Propulsion Laboratory/California Institute of Technology, funded by the National Aeronautics and Space Administration. Scan IDs: 05353b, 05357b, 05361b, 05365b, 05369b, 05372a, 05373b, 05376a, 05377b, 05380a, 05381b, 05384a, 05388a, 05392a, 05396a, 05400a. 
This work is based in part on observations made with the \textit{Spitzer} Space Telescope, which was operated by the Jet Propulsion Laboratory, California Institute of Technology under a contract with NASA. Program ID: 12043 (PI: Mommert). 
This research used the facilities of the Canadian Astronomy Data Centre operated by the National Research Council of Canada with the support of the Canadian Space Agency.
This research has made use of the NASA/IPAC Infrared Science Archive, which is funded by the National Aeronautics and Space Administration and operated by the California Institute of Technology.
This publication makes use of data products from the Near-Earth Object Wide-field Infrared Survey Explorer (NEOWISE), which is a joint project of the Jet Propulsion Laboratory/California Institute of Technology and the University of Arizona. NEOWISE is funded by the National Aeronautics and Space Administration.
This research has made use of the software IRAF. IRAF is distributed by the National Optical Astronomy Observatories, which is operated by the Association of Universities for Research in Astronomy, Inc. (AURA) under cooperative agreement with the National Science Foundation.
This research has made use of NASA’s Astrophysics Data System Bibliographic Services, JPL/Horizons ephemerides service, SAOImageDS9 application, and Python programming language.
\end{acknowledgements}

\bibliographystyle{aa} 
\bibliography{biblio}

\begin{thebibliography}{84}
\expandafter\ifx\csname natexlab\endcsname\relax\def\natexlab#1{#1}\fi

\bibitem[{{A'Hearn} {et~al.}(1984){A'Hearn}, {Schleicher}, {Millis}, {Feldman}, \& {Thompson}}]{A'Hearn1984}
{A'Hearn}, M.~F., {Schleicher}, D.~G., {Millis}, R.~L., {Feldman}, P.~D., \& {Thompson}, D.~T. 1984, \aj, 89, 579

\bibitem[{{Bauer} {et~al.}(2012){Bauer}, {Mainzer}, {Grav}, {Walker}, {Masiero}, {Blauvelt}, {McMillan}, {Fern{\'a}ndez}, {Meech}, {Lisse}, {Cutri}, {Dailey}, {Tholen}, {Riesen}, {Urban}, {Khayat}, {Pearman}, {Scotti}, {Kramer}, {Cherry}, {Gautier}, {Gomillion}, {Watkins}, {Wright}, \& {WISE Team}}]{Bauer2012}
{Bauer}, J.~M., {Mainzer}, A.~K., {Grav}, T., {et~al.} 2012, \apj, 747, 49

\bibitem[{Bodewits {et~al.}(2011)Bodewits, Kelley, Li, Landsman, Besse, \& A’Hearn}]{Bodewits2011}
Bodewits, D., Kelley, M.~S., Li, J.-Y., {et~al.} 2011, ApJL, 733, L3

\bibitem[{Bowell {et~al.}(1989)Bowell, Hapke, Domingue, Lumme, Peltoniemi, \& Harris}]{Bowell1989}
Bowell, E., Hapke, B., Domingue, D., {et~al.} 1989, in Unknown host publication, ed. R.~P. Binzel, T.~Gehrels, \& M.~S. Matthews, 524--556

\bibitem[{{Capria} {et~al.}(2012){Capria}, {Marchi}, {De Sanctis}, {Coradini}, \& {Ammannito}}]{Capria2012}
{Capria}, M.~T., {Marchi}, S., {De Sanctis}, M.~C., {Coradini}, A., \& {Ammannito}, E. 2012, A\&A, 537, A71

\bibitem[{{Cutri} {et~al.}(2012){Cutri}, {Wright}, {Conrow}, {Bauer}, {Benford}, {Brandenburg}, {Dailey}, {Eisenhardt}, {Evans}, {Fajardo-Acosta}, {Fowler}, {Gelino}, {Grillmair}, {Harbut}, {Hoffman}, {Jarrett}, {Kirkpatrick}, {Leisawitz}, {Liu}, {Mainzer}, {Marsh}, {Masci}, {McCallon}, {Padgett}, {Ressler}, {Royer}, {Skrutskie}, {Stanford}, {Wyatt}, {Tholen}, {Tsai}, {Wachter}, {Wheelock}, {Yan}, {Alles}, {Beck}, {Grav}, {Masiero}, {McCollum}, {McGehee}, {Papin}, \& {Wittman}}]{Cutri2012}
{Cutri}, R.~M., {Wright}, E.~L., {Conrow}, T., {et~al.} 2012, {Explanatory Supplement to the WISE All-Sky Data Release Products}, Explanatory Supplement to the WISE All-Sky Data Release Products

\bibitem[{{Divine}(1981)}]{Divine1981}
{Divine}, N. 1981, in ESA Special Publication, Vol. 174, The Comet Halley. Dust and Gas Environment, ed. B.~{Battrick} \& E.~{Swallow}, 47--53

\bibitem[{Drahus {et~al.}(2015)Drahus, Waniak, Tendulkar, Agarwal, Jewitt, \& Sheppard}]{Drahus2015}
Drahus, M., Waniak, W., Tendulkar, S., {et~al.} 2015, ApJL, 802, L8

\bibitem[{Fanale \& Salvail(1989)}]{Fanale1989}
Fanale, F.~P. \& Salvail, J.~R. 1989, Icarus, 82, 97

\bibitem[{{Fazio} {et~al.}(2004){Fazio}, {Hora}, {Allen}, {Ashby}, {Barmby}, {Deutsch}, {Huang}, {Kleiner}, {Marengo}, {Megeath}, {Melnick}, {Pahre}, {Patten}, {Polizotti}, {Smith}, {Taylor}, {Wang}, {Willner}, {Hoffmann}, {Pipher}, {Forrest}, {McMurty}, {McCreight}, {McKelvey}, {McMurray}, {Koch}, {Moseley}, {Arendt}, {Mentzell}, {Marx}, {Losch}, {Mayman}, {Eichhorn}, {Krebs}, {Jhabvala}, {Gezari}, {Fixsen}, {Flores}, {Shakoorzadeh}, {Jungo}, {Hakun}, {Workman}, {Karpati}, {Kichak}, {Whitley}, {Mann}, {Tollestrup}, {Eisenhardt}, {Stern}, {Gorjian}, {Bhattacharya}, {Carey}, {Nelson}, {Glaccum}, {Lacy}, {Lowrance}, {Laine}, {Reach}, {Stauffer}, {Surace}, {Wilson}, {Wright}, {Hoffman}, {Domingo}, \& {Cohen}}]{Fazio2004}
{Fazio}, G.~G., {Hora}, J.~L., {Allen}, L.~E., {et~al.} 2004, \apjs, 154, 10

\bibitem[{{Fink} \& {Rubin}(2012)}]{Fink2012}
{Fink}, U. \& {Rubin}, M. 2012, \icarus, 221, 721

\bibitem[{Gehrz \& Ney(1992)}]{GehrzNey1992}
Gehrz, R. \& Ney, E. 1992, Icarus, 100, 162

\bibitem[{Gwyn {et~al.}(2012)Gwyn, Hill, \& Kavelaars}]{Gwyn2012}
Gwyn, S. D.~J., Hill, N., \& Kavelaars, J.~J. 2012, Publ. Astron. Soc. Pac., 124, 579

\bibitem[{Haghighipour(2009)}]{Haghighipour2009}
Haghighipour, N. 2009, Proceedings of the International Astronomical Union, 5, 207–214

\bibitem[{{Haghighipour} {et~al.}(2016){Haghighipour}, {Maindl}, {Sch{\"a}fer}, {Speith}, \& {Dvorak}}]{Haghighipour2016}
{Haghighipour}, N., {Maindl}, T.~I., {Sch{\"a}fer}, C., {Speith}, R., \& {Dvorak}, R. 2016, \apj, 830, 22

\bibitem[{{Haghighipour} {et~al.}(2018){Haghighipour}, {Maindl}, {Sch{\"a}fer}, \& {Wandel}}]{Haghighipour2018}
{Haghighipour}, N., {Maindl}, T.~I., {Sch{\"a}fer}, C.~M., \& {Wandel}, O.~J. 2018, \apj, 855, 60

\bibitem[{{Hanner} {et~al.}(1981){Hanner}, {Giese}, {Weiss}, \& {Zerull}}]{Hanner1981}
{Hanner}, M.~S., {Giese}, R.~H., {Weiss}, K., \& {Zerull}, R. 1981, \aap, 104, 42

\bibitem[{Holmberg {et~al.}(2006)Holmberg, Flynn, \& Portinari}]{Holmberg2006}
Holmberg, J., Flynn, C., \& Portinari, L. 2006, MNRAS, 367, 449

\bibitem[{Hsieh(2014)}]{Hsieh2014}
Hsieh, H.~H. 2014, Icarus, 243, 16

\bibitem[{Hsieh(2015)}]{HH2015}
Hsieh, H.~H. 2015, Proceedings of the International Astronomical Union, 10, 99–110

\bibitem[{Hsieh {et~al.}(2015)Hsieh, Denneau, Wainscoat, Schörghofer, Bolin, Fitzsimmons, Jedicke, Kleyna, Micheli, Vereš, Kaiser, Chambers, Burgett, Flewelling, Hodapp, Magnier, Morgan, Price, Tonry, \& Waters}]{Hsieh2015}
Hsieh, H.~H., Denneau, L., Wainscoat, R.~J., {et~al.} 2015, Icarus, 248, 289

\bibitem[{Hsieh \& Haghighipour(2016)}]{Hsieh2016}
Hsieh, H.~H. \& Haghighipour, N. 2016, Icarus, 277, 19

\bibitem[{Hsieh {et~al.}(2018{\natexlab{a}})Hsieh, Ishiguro, Kim, Knight, Lin, Micheli, Moskovitz, Sheppard, Thirouin, \& Trujillo}]{Hsieh2018a}
Hsieh, H.~H., Ishiguro, M., Kim, Y., {et~al.} 2018{\natexlab{a}}, AJ, 156, 223

\bibitem[{Hsieh {et~al.}(2011)Hsieh, Ishiguro, Lacerda, \& Jewitt}]{Hsieh2011}
Hsieh, H.~H., Ishiguro, M., Lacerda, P., \& Jewitt, D. 2011, AJ, 142, 29

\bibitem[{Hsieh \& Jewitt(2006)}]{Hsieh2006}
Hsieh, H.~H. \& Jewitt, D. 2006, Science, 312, 561

\bibitem[{{Hsieh} {et~al.}(2009){Hsieh}, {Jewitt}, \& {Fern{\'a}ndez}}]{albedo}
{Hsieh}, H.~H., {Jewitt}, D., \& {Fern{\'a}ndez}, Y.~R. 2009, \apjl, 694, L111

\bibitem[{Hsieh {et~al.}(2008)Hsieh, Jewitt, \& Ishiguro}]{Hsieh2008}
Hsieh, H.~H., Jewitt, D., \& Ishiguro, M. 2008, AJ, 137, 157

\bibitem[{Hsieh {et~al.}(2010)Hsieh, Jewitt, Lacerda, Lowry, \& Snodgrass}]{Hsieh2010}
Hsieh, H.~H., Jewitt, D., Lacerda, P., Lowry, S.~C., \& Snodgrass, C. 2010, MNRAS, 403, 363

\bibitem[{Hsieh {et~al.}(2004)Hsieh, Jewitt, \& Fernández}]{Hsieh2004}
Hsieh, H.~H., Jewitt, D.~C., \& Fernández, Y.~R. 2004, AJ, 127, 2997

\bibitem[{Hsieh {et~al.}(2023)Hsieh, Micheli, Kelley, Knight, Moskovitz, Pittichová, Sheppard, Thirouin, Trujillo, Wainscoat, Weryk, \& Ye}]{Hsieh2023}
Hsieh, H.~H., Micheli, M., Kelley, M. S.~P., {et~al.} 2023, Planet. Sci., 4, 43

\bibitem[{Hsieh {et~al.}(2018{\natexlab{b}})Hsieh, Novaković, Kim, \& Brasser}]{Hsieh2018b}
Hsieh, H.~H., Novaković, B., Kim, Y., \& Brasser, R. 2018{\natexlab{b}}, AJ, 155, 96

\bibitem[{Hsieh \& Sheppard(2015)}]{HsiehSheppard2015}
Hsieh, H.~H. \& Sheppard, S.~S. 2015, MNRASL, 454, L81

\bibitem[{Hsieh {et~al.}(2012{\natexlab{a}})Hsieh, Yang, Haghighipour, Kaluna, Fitzsimmons, Denneau, Novakovi{\'{c} }, Jedicke, Wainscoat, Armstrong, Duddy, Lowry, Trujillo, Micheli, Keane, Urban, Riesen, Meech, Abe, Cheng, Chen, Granvik, Grav, Ip, Kinoshita, Kleyna, Lacerda, Lister, Milani, Tholen, Vere{\v{s}}, Lisse, Kelley, Fern{\'{a}}ndez, Bhatt, Sahu, Kaiser, Chambers, Hodapp, Magnier, Price, \& Tonry}]{Hsieh2012a}
Hsieh, H.~H., Yang, B., Haghighipour, N., {et~al.} 2012{\natexlab{a}}, ApJ, 748, L15

\bibitem[{Hsieh {et~al.}(2012{\natexlab{b}})Hsieh, Yang, Haghighipour, Novakovi{\'{c}}, Jedicke, Wainscoat, Denneau, Abe, Chen, Fitzsimmons, Granvik, Grav, Ip, Kaluna, Kinoshita, Kleyna, Knight, Lacerda, Lisse, Maclennan, Meech, Micheli, Milani, Pittichov{\'{a}}, Schunova, Tholen, Wasserman, Burgett, Chambers, Heasley, Kaiser, Magnier, Morgan, Price, J{\o}rgensen, Dominik, Hinse, Sahu, \& Snodgrass}]{Hsieh2012b}
Hsieh, H.~H., Yang, B., Haghighipour, N., {et~al.} 2012{\natexlab{b}}, AJ, 143, 104

\bibitem[{{Hui} \& {Jewitt}(2017)}]{Hui2017}
{Hui}, M.-T. \& {Jewitt}, D. 2017, \aj, 153, 80

\bibitem[{{IRAC Instrument and Instrument Support Teams}(2021)}]{IRAC_handbook}
{IRAC Instrument and Instrument Support Teams}. 2021, {IRAC Instrument Handbook}, data type: Text

\bibitem[{{Irwin} \& {Lewis}(2001)}]{Irwin2001}
{Irwin}, M. \& {Lewis}, J. 2001, \nar, 45, 105

\bibitem[{Ishiguro {et~al.}(2011)Ishiguro, Hanayama, Hasegawa, Sarugaku, ichi Watanabe, Fujiwara, Terada, Hsieh, Vaubaillon, Kawai, Yanagisawa, Kuroda, Miyaji, Fukushima, Ohta, Hamanowa, Kim, Pyo, \& Nakamura}]{Ishiguro2011}
Ishiguro, M., Hanayama, H., Hasegawa, S., {et~al.} 2011, ApJL, 741, L24

\bibitem[{{Jewitt}(1992)}]{Jewitt1992}
{Jewitt}, D. 1992, Observations and Physical Properties of Small Solar System Bodies, Proceedings of the Liege International Astrophysical Colloquium, 30, 85

\bibitem[{{Jewitt}(1996)}]{Jewitt1996}
{Jewitt}, D. 1996, Earth Moon and Planets, 72, 185

\bibitem[{Jewitt {et~al.}(2014{\natexlab{a}})Jewitt, Agarwal, Li, Weaver, Mutchler, \& Larson}]{Jewitt2014a}
Jewitt, D., Agarwal, J., Li, J., {et~al.} 2014{\natexlab{a}}, ApJL, 784, L8

\bibitem[{Jewitt {et~al.}(2013)Jewitt, Agarwal, Weaver, Mutchler, \& Larson}]{Jewitt2013}
Jewitt, D., Agarwal, J., Weaver, H., Mutchler, M., \& Larson, S. 2013, ApJL, 778, L21

\bibitem[{Jewitt {et~al.}(2016)Jewitt, Agarwal, Weaver, Mutchler, Li, \& Larson}]{Jewitt2016}
Jewitt, D., Agarwal, J., Weaver, H., {et~al.} 2016, AJ, 152, 77

\bibitem[{Jewitt {et~al.}(2015)Jewitt, Hsieh, \& Agarwal}]{Jewitt2015}
Jewitt, D., Hsieh, H., \& Agarwal, J. 2015, in Asteroids {IV} (University of Arizona Press)

\bibitem[{Jewitt {et~al.}(2014{\natexlab{b}})Jewitt, Ishiguro, Weaver, Agarwal, Mutchler, \& Larson}]{Jewitt2014b}
Jewitt, D., Ishiguro, M., Weaver, H., {et~al.} 2014{\natexlab{b}}, AJ, 147, 117

\bibitem[{{Jewitt} \& {Meech}(1988)}]{JewittMeech1988}
{Jewitt}, D. \& {Meech}, K.~J. 1988, \aj, 96, 1723

\bibitem[{Jewitt {et~al.}(2011)Jewitt, Weaver, Mutchler, Larson, \& Agarwal}]{Jewitt2011}
Jewitt, D., Weaver, H., Mutchler, M., Larson, S., \& Agarwal, J. 2011, ApJL, 733, L4

\bibitem[{Jewitt {et~al.}(2009)Jewitt, Yang, \& Haghighipour}]{Jewitt2009}
Jewitt, D., Yang, B., \& Haghighipour, N. 2009, AJ, 137, 4313

\bibitem[{{J}uan {L}uis~{C}ano {R}odr\'{\i}guez \& {J}orge~{M}art\'{\i}nez {G}arrido(2022)}]{poliastro}
{J}uan {L}uis~{C}ano {R}odr\'{\i}guez \& {J}orge~{M}art\'{\i}nez {G}arrido. 2022, in {P}roceedings of the 21st {P}ython in {S}cience {C}onference, ed. {M}eghann {A}garwal, {C}hris {C}alloway, {D}illon {N}iederhut, \& {D}avid {S}hupe, 136 -- 146

\bibitem[{Kelley {et~al.}(2023)Kelley, Hsieh, Bodewits, Saki, Villanueva, Milam, \& Hammel}]{Kelley2023}
Kelley, M. S.~P., Hsieh, H.~H., Bodewits, D., {et~al.} 2023, Nature, 619, 720

\bibitem[{{Kim} {et~al.}(2022){Kim}, {Agarwal}, {Jewitt}, {Mutchler}, {Larson}, {Weaver}, \& {Mommert}}]{kim-agarwal2022}
{Kim}, Y., {Agarwal}, J., {Jewitt}, D., {et~al.} 2022, A\&A, 666, A163

\bibitem[{{Kim} {et~al.}(2017{\natexlab{a}}){Kim}, {Ishiguro}, \& {Lee}}]{Kim2017a}
{Kim}, Y., {Ishiguro}, M., \& {Lee}, M.~G. 2017{\natexlab{a}}, \apjl, 842, L23

\bibitem[{{Kim} {et~al.}(2017{\natexlab{b}}){Kim}, {Ishiguro}, {Michikami}, \& {Nakamura}}]{Kim2017b}
{Kim}, Y., {Ishiguro}, M., {Michikami}, T., \& {Nakamura}, A.~M. 2017{\natexlab{b}}, \aj, 153, 228

\bibitem[{Kolokolova {et~al.}(2004)Kolokolova, Hanner, Levasseur-Regourd, Gustafson, \& Binzel}]{Kolokolova2004}
Kolokolova, L., Hanner, M.~S., Levasseur-Regourd, A.-C., Gustafson, B. A.~S., \& Binzel, R.~P. 2004, Physical Properties of Cometary Dust from Light Scattering and Thermal Emission (University of Arizona Press), 577--604

\bibitem[{{K{\"u}ppers} {et~al.}(2014){K{\"u}ppers}, {O'Rourke}, {Bockel{\'e}e-Morvan}, {Zakharov}, {Lee}, {von Allmen}, {Carry}, {Teyssier}, {Marston}, {M{\"u}ller}, {Crovisier}, {Barucci}, \& {Moreno}}]{Kuppers2014}
{K{\"u}ppers}, M., {O'Rourke}, L., {Bockel{\'e}e-Morvan}, D., {et~al.} 2014, \nat, 505, 525

\bibitem[{{Lagerkvist} \& {Magnusson}(1990)}]{Lagerkvist1990}
{Lagerkvist}, C.~I. \& {Magnusson}, P. 1990, \aaps, 86, 119

\bibitem[{{Magnier}(2006)}]{Magnier2006}
{Magnier}, E. 2006, in The Advanced Maui Optical and Space Surveillance Technologies Conference, E50

\bibitem[{{Magnier} \& {Cuillandre}(2004)}]{Magnier2004}
{Magnier}, E.~A. \& {Cuillandre}, J.~C. 2004, \pasp, 116, 449

\bibitem[{{Mainzer} {et~al.}(2011){Mainzer}, {Grav}, {Masiero}, {Bauer}, {Wright}, {Cutri}, {McMillan}, {Cohen}, {Ressler}, \& {Eisenhardt}}]{Mainzer2011}
{Mainzer}, A., {Grav}, T., {Masiero}, J., {et~al.} 2011, \apj, 736, 100

\bibitem[{Marinelli \& Dressel(2024)}]{WFCHandbook}
Marinelli, M. \& Dressel, L. 2024, Wide Field Camera 3 Instrument Handbook, Version 16.0 (Baltimore, MD: Space Telescope Science Institute)

\bibitem[{{Masiero} {et~al.}(2011){Masiero}, {Mainzer}, {Grav}, {Bauer}, {Cutri}, {Dailey}, {Eisenhardt}, {McMillan}, {Spahr}, {Skrutskie}, {Tholen}, {Walker}, {Wright}, {DeBaun}, {Elsbury}, {Gautier}, {Gomillion}, \& {Wilkins}}]{Masiero2011}
{Masiero}, J.~R., {Mainzer}, A.~K., {Grav}, T., {et~al.} 2011, \apj, 741, 68

\bibitem[{Moreno {et~al.}(2013)Moreno, Cabrera-Lavers, Vaduvescu, Licandro, \& Pozuelos}]{Moreno2013}
Moreno, F., Cabrera-Lavers, A., Vaduvescu, O., Licandro, J., \& Pozuelos, F. 2013, ApJL, 770, L30

\bibitem[{Moreno {et~al.}(2011)Moreno, Lara, Licandro, Ortiz, de~León, Alí-Lagoa, Agís-González, \& Molina}]{Moreno2011}
Moreno, F., Lara, L.~M., Licandro, J., {et~al.} 2011, ApJL, 738, L16

\bibitem[{{Nomen} {et~al.}(2010){Nomen}, {Birtwhistle}, {Holmes}, {Foglia}, \& {Scotti}}]{Nomen2010}
{Nomen}, J., {Birtwhistle}, P., {Holmes}, R., {Foglia}, S., \& {Scotti}, J.~V. 2010, Central Bureau Electronic Telegrams, 2459, 1

\bibitem[{{Ootsubo} {et~al.}(2011){Ootsubo}, {Kawakita}, {Hamada}, {Kobayashi}, {Yamaguchi}, \& {Usui}}]{Ootsubo2011}
{Ootsubo}, T., {Kawakita}, H., {Hamada}, S., {et~al.} 2011, in EPSC-DPS Joint Meeting 2011, Vol. 2011, 369

\bibitem[{{Polishook} \& {Brosch}(2009)}]{Polishook2009}
{Polishook}, D. \& {Brosch}, N. 2009, \icarus, 199, 319

\bibitem[{Pozuelos {et~al.}(2015)Pozuelos, Cabrera-Lavers, Licandro, \& Moreno}]{Pozuelos2015}
Pozuelos, F.~J., Cabrera-Lavers, A., Licandro, J., \& Moreno, F. 2015, ApJ, 806, 102

\bibitem[{Prialnik \& Rosenberg(2009)}]{Prialnik2009}
Prialnik, D. \& Rosenberg, E.~D. 2009, MNRASL, 399, L79

\bibitem[{{Reach} {et~al.}(2013){Reach}, {Kelley}, \& {Vaubaillon}}]{reach-kelley2013}
{Reach}, W.~T., {Kelley}, M.~S., \& {Vaubaillon}, J. 2013, Icarus, 226, 777

\bibitem[{{Russell}(1916)}]{Russell1916}
{Russell}, H.~N. 1916, \apj, 43, 173

\bibitem[{Sarmecanic {et~al.}(1997)Sarmecanic, Fomenkova, Jones, \& Lavezzi}]{Sarmecanic1997}
Sarmecanic, J., Fomenkova, M., Jones, B., \& Lavezzi, T. 1997, ApJ, 483, L69

\bibitem[{{Schleicher}(2010)}]{lowell_dustphase}
{Schleicher}, D. 2010, Composite Dust Phase Function for Comets, \url{https://asteroid.lowell.edu/comet/dustphase/}, Accessed: July 3, 2024

\bibitem[{Schorghofer(2008)}]{Schorghofer2008}
Schorghofer, N. 2008, ApJ, 682, 697

\bibitem[{Schorghofer(2016)}]{Schorghofer2016}
Schorghofer, N. 2016, Icarus, 276, 88

\bibitem[{{Sch{\"o}rghofer} \& {Hsieh}(2018)}]{SchorghoferHsieh2018}
{Sch{\"o}rghofer}, N. \& {Hsieh}, H.~H. 2018, Journal of Geophysical Research (Planets), 123, 2322

\bibitem[{Sheppard \& Trujillo(2015)}]{Sheppard2015}
Sheppard, S.~S. \& Trujillo, C. 2015, AJ, 149, 44

\bibitem[{{Thiel} {et~al.}(1989){Thiel}, {Koelzer}, {Kochan}, {Ratke}, {Gruen}, \& {Koehl}}]{Thiel1989}
{Thiel}, K., {Koelzer}, G., {Kochan}, H., {et~al.} 1989, In ESA, Physics and Mechanics of Cometary Materials, 302, 221

\bibitem[{{Tody}(1986)}]{iraf1}
{Tody}, D. 1986, in Society of Photo-Optical Instrumentation Engineers (SPIE) Conference Series, Vol. 627, Instrumentation in astronomy VI, ed. D.~L. {Crawford}, 733

\bibitem[{{Tody}(1993)}]{iraf2}
{Tody}, D. 1993, in Astronomical Society of the Pacific Conference Series, Vol.~52, Astronomical Data Analysis Software and Systems II, ed. R.~J. {Hanisch}, R.~J.~V. {Brissenden}, \& J.~{Barnes}, 173

\bibitem[{{Tonry} {et~al.}(2012){Tonry}, {Stubbs}, {Lykke}, {Doherty}, {Shivvers}, {Burgett}, {Chambers}, {Hodapp}, {Kaiser}, {Kudritzki}, {Magnier}, {Morgan}, {Price}, \& {Wainscoat}}]{Tonry2012}
{Tonry}, J.~L., {Stubbs}, C.~W., {Lykke}, K.~R., {et~al.} 2012, \apj, 750, 99

\bibitem[{Williams {et~al.}(1997)Williams, Mason, Gehrz, Jones, Woodward, Harker, Hanner, Wooden, Witteborn, \& Butner}]{Williams1997}
Williams, D.~M., Mason, C.~G., Gehrz, R.~D., {et~al.} 1997, ApJ, 489, L91

\bibitem[{{Willmer}(2018)}]{Willmer2018}
{Willmer}, C. N.~A. 2018, \apjs, 236, 47

\bibitem[{Wright {et~al.}(2010)Wright, Eisenhardt, Mainzer, Ressler, Cutri, Jarrett, Kirkpatrick, Padgett, McMillan, Skrutskie, Stanford, Cohen, Walker, Mather, Leisawitz, Gautier, McLean, Benford, Lonsdale, Blain, Mendez, Irace, Duval, Liu, Royer, Heinrichsen, Howard, Shannon, Kendall, Walsh, Larsen, Cardon, Schick, Schwalm, Abid, Fabinsky, Naes, \& Tsai}]{Wright2010}
Wright, E.~L., Eisenhardt, P. R.~M., Mainzer, A.~K., {et~al.} 2010, AJ, 140, 1868

\bibitem[{Yang {et~al.}(2009)Yang, Jewitt, \& Bus}]{Yang2009}
Yang, B., Jewitt, D., \& Bus, S.~J. 2009, AJ, 137, 4538

\end{thebibliography}

\begin{appendix}
\section{Figure and table} 

\begin{figure*}[b]
\centering
\includegraphics[width=17cm]{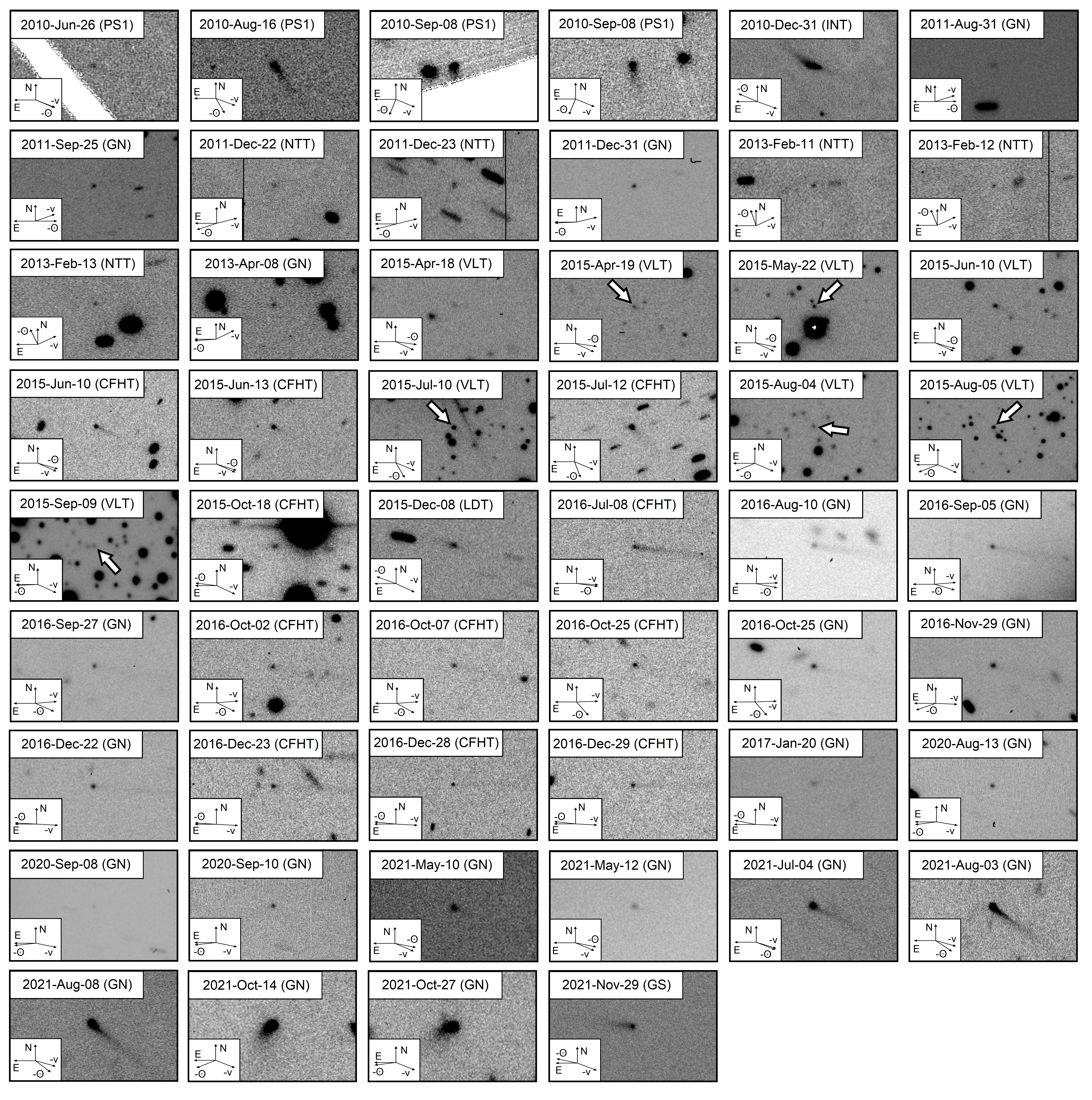}
\caption{Mosaic of the visible-light images of 324P (at the center of each panel). All images are in the R band, except for two PS1 images on 2010 June 26 in the z band, and on 2010 September 8 in the g band. 
All panels show the north ($N$), the east ($E$), the antisolar direction ($-\odot$) and the negative heliocentric velocity vector ($-v$), as projected on the sky. 
Panels are labeled with dates of observation in UT YYYY-MM-DD format, and employed telescope.
The label PS1 refers to the Panoramic Survey Telescope and Rapid Response System PanSTARRS (PS1) telescope, equipped with the GPC1 instrument; INT to the Isaac Newton Telescope, equipped with the WFC instrument; GN to the Gemini Multi-Object Spectrograph North, equipped with the Gemini Multi-Object Spectrograph North (GMOS-N) instrument; NTT to the New Technology Telescope, equipped with the ESO Faint Object Spectrograph and Camera (EFOSC) instrument; VLT to the Very Large Telescope, equipped with the FOcal Reducer and low dispersion Spectrograph 2 (FORS2) instrument; CFHT to the Canada-France-Hawaii Telescope, equipped with the MegaPrime instrument; LDT to the Lowell Discovery Telescope, equipped with the Large Monolithic Images (LMI) instrument; and GS to the Gemini Multi-Object Spectrograph South, equipped with the Gemini Multi-Object Spectrograph South (GMOS-S) instrument.}
\label{mosaic}
\end{figure*}

In this appendix, Fig~\ref{mosaic} and Table~\ref{table:1} are provided.

\begin{table*}[h]
\centering
\caption{Observations of the MBC 324P.} 
\label{table:1}
\begin{threeparttable}
\begingroup
\setlength{\tabcolsep}{2.5pt} 
\renewcommand{\arraystretch}{.4} 
\begin{tabular}{lcccrrrrrrr}
\toprule
UT Date & Telescope/Instrument$^a$ & Pixel scale$^b$ &  N$^c$& $r_h$$^d$ & $\mathit{\Delta}$$^e$ & $\alpha$$^f$ & $\nu$$^g$  & $m_{R}$$^h$ & $H_{R}$$^i$  & $M_{d}$ (x $10^6$)$^l$ \\
\midrule
2010 Jun 25 & \textit{Perihelion} &--& --&   2.623  & 2.239 & 22.4 &  0.0  & --& --& -- \\
2010 Jun 26 & PS1/GPC1 &0.25  &1 &  2.623   &  2.228&  22.3  &    0.0      &20.7 $\pm$ 0.5&15.8 $\pm$ 0.5&28 $\pm$ 18 \\ 
2010 Aug 16* & PS1/GPC1 &0.25& 4 & 2.631& 1.795 & 15.1  & 12.8 & 19.19 $\pm$ 0.02 &  15.0 $\pm$ 0.1 & 63 $\pm$ 26\\ 
2010 Sep 08* & PS1/GPC1 &0.25&1 &    2.641&  1.739  &	12.0 & 18.4    &19.14 $\pm$ 0.05 &15.1 $\pm$ 0.1&56 $\pm$ 23 \\ 
2010 Sep 08 & PS1/GPC1 &0.25&4   &  2.641&  1.739  &	12.0 & 18.4   &18.9 $\pm$ 0.1&14.9 $\pm$ 0.2&70 $\pm$ 31  \\ 
2010 Dec 31* &  INT/WFC &0.33&   11 &     2.732  & 2.782  & 20.5  & 45.7        &20.8 $\pm$ 0.1&15.4 $\pm$ 0.2& 42 $\pm$ 19\\ 
2011 Aug 31* & GN/GMOS-N &0.14&  6 &   3.072 & 3.238 &	18.2  & 95.9   &24.4 $\pm$ 0.1&18.5 $\pm$ 0.2& -0.2 $\pm$ 1.4\\ 
2011 Sep 25**& GN/GMOS-N &0.14& 9 &     3.110 &  2.929 	&18.8 & 100.4    &23.8 $\pm$ 0.1&18.0 $\pm$ 0.2& 1.2 $\pm$ 1.6\\ 
2011 Dec 22   &NTT/EFOSC  &0.24& 8   &  3.236 &  2.257   &1.9 & 115.5&    23.0 $\pm$ 0.2 & 18.4 $\pm$ 0.2&     0.0 $\pm$ 1.4    \\ 
2011 Dec 23   &NTT/EFOSC  & 0.24 & 10   & 3.238  &  2.260   & 2.2  &115.7&  22.9 $\pm$ 0.2 & 18.3 $\pm$ 0.2&    0.3 $\pm$ 1.4    \\ 
2011 Dec 31** &GN/GMOS-N &0.14& 9&    3.249 &  2.295   &  5.0 &  117.0   &23.0 $\pm$ 0.1 &18.2 $\pm$ 0.1&0.6 $\pm$ 1.4\\ 
2013 Feb 11   & NTT/EFOSC & 0.24 &  27  & 3.568  &   2.643    &   6.4 & 176.0&   24.0 $\pm$ 0.2    &18.6 $\pm$ 0.2&-0.5 $\pm$ 1.4      \\ 
2013 Feb 12   & NTT/EFOSC & 0.24 &  10  & 3.568  &   2.643  & 6.4  &176.2&   23.7 $\pm$ 0.2&18.3 $\pm$ 0.2&    0.3 $\pm$ 1.4\\ 
2013 Feb 13  & NTT/EFOSC  & 0.24 &  9  & 3.568  &   2.643  &  6.4&  176.3       & 23.9 $\pm$ 0.1&18.5 $\pm$ 0.1&    -0.2 $\pm$ 1.3\\ 
2013 Mar 12 & \textit{Aphelion} &-- & --&   3.569  & 2.750 & 10.3 &  180.0  & --& --& -- \\
2013 Apr 08**& GN/GMOS-N  &0.14& 8 &         3.568  &  3.017&14.6 & 183.6&24.2 $\pm$ 0.2&18.2 $\pm$ 0.2 &0.6 $\pm$ 1.5\\ 
2015 Apr 18   & VLT/FORS2  &0.25&   15  & 2.773  &  2.559   &   21.2 & 305.7&     23.3 $\pm$ 0.3&18.0 $\pm$ 0.3&     1.2 $\pm$ 1.8\\ 
2015 Apr 19   &  VLT/FORS2 &0.25 & 18   &  2.772 &   2.544   & 21.2&  306.0&    23.1 $\pm$ 0.1&17.8 $\pm$ 0.2&  2.1 $\pm$ 1.8       \\ 
2015 May 22   & VLT/FORS2  &0.25 &  16  &  2.734 &   2.088  &  18.8& 313.4&    22.6 $\pm$ 0.1    &17.8 $\pm$ 0.2&     2.1 $\pm$ 1.8\\ 
2015 Jun 10   &  VLT/FORS2  &0.25&  18  &2.713   &   1.878  &14.7 & 317.8&    22.2 $\pm$ 0.1    &     17.8 $\pm$ 0.2&2.1 $\pm$ 1.8\\ 
2015 Jun 10***&   CFHT/MegaPrime &0.18& 3 & 2.713  & 1.878  &  14.7  &317.8  & 21.959 $\pm$ 0.002   &17.6 $\pm$ 0.1  &3.1 $\pm$ 1.9	 \\ 
2015 Jun 13*** &CFHT/MegaPrime  &0.18& 9 & 2.710  &  1.850  &  13.8  &  	318.5  &  22.2 $\pm$ 0.2 &   17.9 $\pm$ 0.2 & 1.6 $\pm$ 1.7	  \\ 
2015 Jul 10 &VLT/FORS2 &0.25 & 5  &2.685 &  1.693   &  6.0   & 324.9&21.7  $\pm$ 0.1	&17.9 $\pm$ 0.1&1.6 $\pm$ 1.5	\\ 
2015 Jul 12 &CFHT/MegaPrime  &0.18&  3 &   2.683  &   1.688  &    5.6  &   325.4  &   21.38 $\pm$ 0.04 & 17.6 $\pm$ 0.1  & 	2.8 $\pm$ 1.8  \\ 
2015 Aug 04   & VLT/FORS2 &0.25  & 9   &  2.665 &  1.715   &9.6  &330.9& 22.0 $\pm$ 0.3&     18.1 $\pm$ 0.3&     0.9 $\pm$ 1.7     \\ 
2015 Aug 05 & VLT/FORS2 &0.25 &14 &	   2.664&   1.719 &    9.9 &   331.1 &21.8 $\pm$ 0.1&	    17.8 $\pm$ 0.1	&	   2.1 $\pm$ 1.6	\\ 
2015 Sep 09  &VLT/FORS2 &0.25&   4&    2.642    &  1.990   &  19.3    & 339.6    &22.4 $\pm$ 0.1   &17.8 $\pm$ 0.2   &2.1 $\pm$ 1.8\\ 
2015 Oct 18    &CFHT/MegaPrime &0.18& 2 &   2.626    & 2.443    & 22.3   &  349.3   &22.6 $\pm$ 0.1   &    17.5 $\pm$ 0.2   &3.6 $\pm$ 2.3\\ 
2015 Nov 30 & \textit{Perihelion} & --& --&   2.620  & 2.946 & 19.3 &  0.0  & --& --& -- \\
2015 Dec 08 &LDT/LMI &0.24& 10&      2.620& 3.029  &18.3  &2.0 & 22.2 $\pm$ 0.1 & 16.8 $\pm$ 0.2 &9.5 $\pm$ 4.6 \\ 
2016 Jul 08 & CFHT/MegaPrime & 0.18& 6 &    2.768  &   3.067  &    19.2  &    53.2  &  22.8 $\pm$ 0.1 & 17.2 $\pm$ 0.2  &5.7 $\pm$ 3.1  \\ 
2016 Aug 10& GN/GMOS-N &0.14& 4 & 2.810    &  2.724   &  21.0  &    60.5&23.1 $\pm$ 0.1&17.6 $\pm$ 0.2&   3.1 $\pm$ 2.1   \\ 
2016 Sep 05&GN/GMOS-N &0.14&4  &  2.845    & 2.436    & 20.2     & 66.0   &   23.3 $\pm$ 0.1&18.1 $\pm$ 0.2   &0.9 $\pm$ 1.5 \\ 
2016 Sep 27 &GN/GMOS-N &0.14& 5 & 2.876  &   2.214  &  17.2   &70.7&23.0 $\pm$ 0.1   &		18.0 $\pm$ 0.2 	 &	1.2 $\pm$ 1.5	\\ 
2016 Oct 02 & CFHT/MegaPrime &0.18 &5  &2.883  &   2.171  &    16.2  &    71.7  &   22.7 $\pm$ 0.1 & 17.8 $\pm$ 0.2  & 	  2.1 $\pm$ 1.8	  \\ 
2016 Oct 07 &CFHT/MegaPrime &0.18 &  8&   2.890  &   2.131   &   15.0  &    72.7  &  22.9 $\pm$ 0.2 & 18.1 $\pm$ 0.2  & 	0.9 $\pm$ 1.5  \\ 
2016 Oct 25&CFHT/MegaPrime &0.18 & 5  &2.917  &   2.026  &    10.4  &   76.4  &   22.5 $\pm$ 0.1 & 18.0 $\pm$ 0.1  & 	1.2 $\pm$ 1.4\\ 
2016 Oct 25&GN/GMOS-N &0.14& 4 &2.917   & 2.026&  10.4    &   76.4&22.54 $\pm$ 0.04   &	18.0 $\pm$ 0.1    &	1.2 $\pm$ 1.4\\ 
2016 Nov 29 &GN/GMOS-N &0.14& 3 &  2.969   &   2.054   &   8.5   &   83.3&	22.7 $\pm$ 0.1&18.2 $\pm$ 0.1	 &	0.7 $\pm$ 1.4\\ 
2016 Dec 22 &GN/GMOS-N &0.14& 4 &3.004 &    2.250    &  13.9   &   87.7&		23.1 $\pm$ 0.1&18.2 $\pm$ 0.2 	 &0.6 $\pm$ 1.5	\\ 
2016 Dec 23 & CFHT/MegaPrime & 0.18&5  &   3.006  &   2.261  &    14.2  &   87.9  &   23.0 $\pm$ 0.1 & 18.0 $\pm$ 0.2 & 	1.2 $\pm$ 1.6  \\ 
2016 Dec 28 &CFHT/MegaPrime &0.18 & 9  &   3.013  &   2.320  &    15.2  &    88.9  &  23.1 $\pm$ 0.3 & 18.0 $\pm$ 0.3  &1.2 $\pm$ 1.8\\ 
2016 Dec 29 &  CFHT/MegaPrime &0.18 & 5&   3.015  &   2.332  &    15.3  &    89.1  &  23.2 $\pm$ 0.1 & 18.1 $\pm$ 0.2  &0.9 $\pm$ 1.5	 \\ 
2017 Jan 20 &GN/GMOS-N &0.14& 5 &  3.048    & 2.636   &  18.1   &   93.2&    23.9 $\pm$ 0.1&18.4 $\pm$ 0.2&	0.0 $\pm$ 1.4	\\ 
2018 Aug 19 & \textit{Aphelion} &--& --&   3.571   &  4.544  &    4.0&   180.0  & --& --& -- \\
2020 Aug 13 &GN/GMOS-N &0.16& 4 & 2.826 &    2.390 &    20.3 &  296.7 &     23.2 $\pm$ 0.2&     18.0 $\pm$ 0.2&     1.2 $\pm$ 1.6       \\ 
2020 Sep 08 &GN/GMOS-N &0.16& 3 & 2.792 &    2.705 &     21.0 &  302.4 &     23.7 $\pm$ 0.2&     18.3 $\pm$ 0.2&     0.3 $\pm$ 1.4       \\ 
2020 Sep 10 &GN/GMOS-N &0.16& 4 &  2.789 &    2.729  &    21.0 & 302.8 &     23.7 $\pm$ 0.1&     18.3 $\pm$ 0.2&     0.3 $\pm$ 1.4      \\ 
2021 May 06 & \textit{Perihelion} & --& --&   2.618  & 2.972 & 19.5 &  0.0  & --& --& -- \\ 
2021 May 10  &  GN/GMOS-N &0.16&    3   &  2.619 &  2.931 & 	20.0 &   1.0 &     23.35 $\pm$ 0.05   &  17.9 $\pm$ 0.2 &   1.5 $\pm$ 1.6    \\ 
2021 May 12  &  GN/GMOS-N &0.16&  3&     2.619 &   2.910 &	20.2 &    1.5    &  23.38 $\pm$ 0.03  &    18.0 $\pm$ 0.2 &    1.4 $\pm$ 1.6   \\ 
2021 Jul 04  &  GN/GMOS-N &0.16&    3&   2.630 &  2.330 &	22.6 & 14.7      &     22.29 $\pm$ 0.06&    17.3 $\pm$ 0.2   &    5.2 $\pm$ 2.8     \\ 
2021 Aug 03  &  GN/GMOS-N &0.16&   3&    2.645 &   2.033 & 	20.1 &   22.2     &21.57 $\pm$ 0.05&    16.9 $\pm$ 0.2     &   8.3 $\pm$ 3.9    \\ 
2021 Aug 08  &  GN/GMOS-N &0.16&   3&    2.648 &   1.991 & 	19.4 &   23.4     &21.51 $\pm$ 0.01&   16.9 $\pm$ 0.2&    8.2 $\pm$ 3.9      \\ 
2021 Oct 14  &  GN/GMOS-N &0.16&   3&       2.701 &  1.839 & 	13.0 &  39.5  &20.82 $\pm$ 0.04&     16.6 $\pm$ 0.1      &    12 $\pm$ 5     \\ 
2021 Oct 27  &  GN/GMOS-N &0.16&   3&   2.714 &  1.920 &	15.0 &   42.5      &21.00 $\pm$ 0.07&     16.6 $\pm$ 0.2       &   12 $\pm$ 6    \\ 
2021 Nov 29  &  GS/GMOS-S &0.16&  4&   2.750 &  2.253 &	19.7 &   50.1      &21.86 $\pm$ 0.04     &    16.9 $\pm$ 0.2       &     8.3 $\pm$ 3.9     \\ 
2024 Jan 24 & \textit{Aphelion} &-- & --&   3.570  &   2.750  &  10.0 &   180.0  & --& --& -- \\
2026 Oct 14 & \textit{Perihelion} & --& --&   2.625 &  2.232 &   21.8 &  0.0  & --& --& -- \\ 
\hline
2010 June 9 -- 11 &  WISE/W1-2-3-4 &2.75, 5.5&19, 18 & 2.623  & 2.410   &22.8 &   356.1  &--&--&--\\
2016 Jan 08	& \textit{Spitzer}/IRAC1-2 &1.22& 70, 71   & 2.625    & 1.964  &     19.8  &  9.8   &--&--& --\\
\hline
2015 Sep 28 &  HST/WFC3 &0.04& 4 &2.633  &  2.202  &  21.6 & 344.3   &22.0 $\pm$ 0.1 &17.1 $\pm$ 0.2 & \\ 
2015 Oct 08 &   HST/WFC3 &0.04& 4 & 2.629  &2.321  &22.2  &346.8   &22.4 $\pm$ 0.1&17.4 $\pm$ 0.2  &--\\ 
2015 Dec 07 & HST/WFC3 &0.04& 4 &2.620   &3.019  &18.4    &1.8    & 22.2 $\pm$ 0.1&16.8 $\pm$ 0.2  &--\\ 
2015 Dec 18 &   HST/WFC3 &0.04& 4 & 2.621   &3.127  &16.9  &  4.5  & 21.9 $\pm$ 0.1&16.4 $\pm$ 0.2  &--\\ 
\bottomrule
\end{tabular}
\begin{tablenotes}\small{
\item$^a$Telescope and instrument. $^b$Pixel scale in arcsec per pixel. $^c$Number of exposures taken. $^d$Heliocentric distance in AU. $^e$Geocentric distance in AU. $^f$Solar phase angle (Sun-Target-Observer) in degrees. $^g$True anomaly in degrees. $^h$Apparent R-band magnitude measured with a 2300 km physical radius. $^i$Absolute R-band magnitude assuming IAU H-G phase function where $G = 0.15 \pm 0.12$. $^l$Estimated dust mass in kg assuming $\rho_{d}$ $\sim$ 2500 kg m$^{-3}$. \\
Notes: Data presented in the last three columns are the results of our own photometric analysis. Data from dates marked with asterisks were previously published in \citet{Hsieh2012b} (*), \citet{Hsieh2014} (**), and \citet{HsiehSheppard2015} (***). The HST dataset has been previously analyzed in \citet{Jewitt2016}.
For WISE, 19 images were obtained in W1 and W2 bands, and 18 in W3 and W4 bands. For \textit{Spitzer}, 70 images were taken in ch1 and 71 in ch2. 
For HST, the apparent V-band magnitude was measured inside a 3000 km physical radius, and the absolute V-band magnitude was derived assuming an IAU H-G phase function with $G = 0.15 \pm 0.12$.
}
\end{tablenotes}
\endgroup
\end{threeparttable}
\end{table*}
\FloatBarrier

\end{appendix}

\end{document}